\documentclass{emulateapj}

\slugcomment{accepted for publication in ApJ}

\shorttitle{Debris Disk Evolution around A Stars}
\shortauthors{Su et al.}

\newcommand{\um}{${\rm \mu m}$~}
\newcommand{\mm}{${\rm \mu m}$}

\begin{document}

\title{Debris Disk Evolution around A Stars}
\author{
  K. Y. L. Su\altaffilmark{1}, 
  G. H. Rieke\altaffilmark{1}, 
  J. A. Stansberry\altaffilmark{1}, 
  G. Bryden\altaffilmark{2}, 
  K. R. Stapelfeldt\altaffilmark{2}, 
  D. E. Trilling\altaffilmark{1},
  J. Muzerolle\altaffilmark{1},
  C. A. Beichman\altaffilmark{2,3}
  A. Moro-Martin\altaffilmark{4},
  D. C. Hines\altaffilmark{5},  
  M. W. Werner\altaffilmark{2} \\
  }

\altaffiltext{1}{Steward Observatory, University of Arizona, 933 N
  Cherry Ave., Tucson, AZ 85721; ksu@as.arizona.edu}
\altaffiltext{2}{JPL/Caltech, 4800 Oak Grove Drive, Pasadena, CA
  91109}
\altaffiltext{3}{Michelson Science Center, California Institute of Technology, M/S 100-22, Pasadena, CA 91125}
\altaffiltext{4}{Department of Astrophysical Sciences, Princeton
  University, Princeton, NJ 08540}
\altaffiltext{5}{Space Science Institute, 4700 Walnut St. Suit 205, Boulder,
  Colorado 80301}

\begin{abstract}

We report 24 and/or 70 \um measurements of $\sim$160 A-type
main-sequence stars using the Multiband Imaging Photometer for {\it
Spitzer} (MIPS). Their ages range from 5 to 850 Myr based on
estimates from the literature (cluster or moving group associations)
or from the H-R diagram and isochrones. The thermal infrared excess is
identified by comparing the deviation ($\sim$3\% and $\sim$15\% at the
1-$\sigma$ level at 24 and 70 \mm, respectively) between the
measurements and the synthetic Kurucz photospheric predictions.  Stars
showing excess infrared emission due to strong emission lines or
extended nebulosity seen at 24 \um are excluded from our sample;
therefore, the remaining infrared excesses are likely to arise from
circumstellar debris disks.  At the 3-$\sigma$ confidence level, the
excess rate at 24 and 70 \um is 32\% and $\ge$33\% (with an uncertainty
of 5\%), considerably higher than has been found for old solar analogs
and M dwarfs. Our measurements place constraints on the fractional
dust luminosities and temperatures in the disks.  We find that older
stars tend to have lower fractional dust luminosity than younger ones.
While the fractional luminosity from the excess infrared emission
follows a general 1/$t$ relationship, the values at a given stellar
age vary by at least two orders of magnitude. We also find that (1)
older stars possess a narrow range of temperature distribution peaking
at colder temperatures, and (2) the disk emission at 70 \um persists
longer than that at 24 \mm. Both results suggest that the debris-disk
clearing process is more effective in the inner regions.

\end{abstract} 


\section{Introduction}

From various lines of evidence (e.g., theoretical modeling, the
cratering record on the Moon, returned Lunar samples, and isotopic
data) it is believed that embryo terrestrial planets in the Solar
System formed in $\sim$10-30 Myr, evolved through a period of
potentially immense collisions, and then underwent a reduced, but
still significant collisional period that ended with another violent
episode -- the Late Heavy Bombardment -- at about 700 Myr
\citep{kleine02,chambers04,gomes05,strom05}. Thereafter, the system
was in a relatively settled state, setting the stage for life to
emerge on Earth. We are unsure whether this sequence was typical or
exceptional. We are not even sure of the details of the steps because
much of the evidence has been obliterated over time. Hard evidence
must often be supplemented with relatively poorly tested theory to
assemble a complete picture. The obvious solution would be to observe
these processes as they are occurring in other planetary
systems. However, observing terrestrial planets around nearby stars
passing through parallel stages of evolution is virtually
impossible. The planets are too dim for detection against the glare of
the central stars; they are not sufficiently massive to be detected
through gravitational recoil; and relatively few young stars lie close
to the Sun.

However, there is a promising alternative approach, to study planetary
debris disks. Debris disks arise from collisions of asteroidal (or
planetesimal) bodies that lead to cascades of collisions among the
resulting debris. Eventually, significant amounts of this material are
ground down to dust grains. Because the surface area per unit mass is
large for dusty material, when these grains are heated by the central
star they can produce a readily detectable level of excess emission in
the mid- and far-infrared. The behavior of these infrared excesses can
trace the different zones within a planetary system. For grains around
an A-type star and large enough to be in pseudo-stable orbits, the
mid-IR band ({\it Spitzer} 24 \mm, {\it IRAS} and {\it ISO} 25 \um
passbands) is sensitive to material largely between $\le$5 and
$\sim$50 AU, while the far-IR band ({\it IRAS} and {\it ISO} 60 \um
and {\it Spitzer} 70 \um passbands) is sensitive to material between
$\sim$50 and 200 AU. Detailed studies of the behavior with age of
debris disks in these two bands can indicate how the stages deduced
for the early evolution of the Solar System are playing out in
hundreds of other planetary systems.

Pioneering studies with {\it IRAS} and {\it ISO} suggest a systematic
drop in infrared excess with stellar age (e.g.,
\citealt{habing01,spangler01}), qualitatively similar to the drop that
would be deduced from the settling down of the Solar System. However,
because of limitations in the sensitivity and the accuracy of
measurements with these two missions, they were unable to show
unambiguously how debris disks evolve. {\it Spitzer} brings
significant advances in both sensitivity and photometric accuracy,
increasing our ability to detect low levels of infrared excesses
around hundreds of stars. A-type stars provide an ideal laboratory to
study the early stages in planetary system evolution, particularly in
the zones relevant for evolution of terrestrial planets (Earth-like
temperatures occur at about 5 AU from such stars). Their main sequence
lifetimes are long enough ($\sim$800 Myr) to encompass the entire
period of interest, they are of high enough luminosity to light up
their debris well, they are sufficiently abundant and bright that many
can be observed readily, and they are cool enough that they are
unlikely to create infrared excesses in the form of ionized
gas. \citet{rieke05} have already conducted a study of the 24/25 \um
excesses in such stars, based on a combination of {\it IRAS} and
Multiband Imaging Photometer for {\it Spitzer} (MIPS) measurements of
stars in the field and MIPS measurements of stars in clusters. At that
time, relatively few field stars had been measured well with {\it
Spitzer} at 70 \mm. In this paper, we report MIPS photometry at 24 and
70 \um of 160 main-sequence early type stars, enough to give a much
better understanding of the evolution of debris systems.

\section{Observations and Data Reduction} 
\label{astar_obs} 

\subsection{Sample Selection}

The majority (128) of the stars in this study are from the {\it
Spitzer} Guaranteed Time Observation (GTO) programs (Astar, Fab4, and
Dirty12), where most of them were selected to have known ages based on
cluster membership or association with moving groups. An additional 32
stars are included from the {\it Spitzer} calibration observations. In
total, our sample includes 160 stars ranging in spectral type from B6
to A7, but mostly ($\sim$85\%) from B9 to A7. We have also developed a
tool to determine ages from the Hertzsprung-Russell diagram (HRD)
\citep{rieke05} for consistency checking and estimating ages for the
stars that are not in clusters or moving group associations. Two stars
in the calibration programs are too bright to be observed at 24 \mm,
while only 9 stars are bright enough at 70 \um to be observed as
calibrators. Hence, a total of 158 MIPS 24 \um measurements and 137
MIPS 70 \um measurements are reported here. The AOR keys, stellar
properties and estimated ages are listed in Table \ref{tab_sample}.

Most of the stars in our sample are single stars (according to the
literature and/or SIMBAD database); there are only 10 binary/multiple
systems. These are either spectroscopic binaries (unresolved in
ground-based optical photometry) or wide binaries with separations
larger than the MIPS 24 \um beam (resolved at 24 \mm). In addition, we
also double checked the stars with infrared excesses (identified by
the method discussed below) in the 2MASS catalog, and found no nearby
objects on the sky that could confuse the MIPS 24 and 70 \um
measurements. Therefore, the false detection of infrared excesses due
to binary components or positional coincidence is unlikely.

\subsection{Data Reduction and Source Extraction}

All data were processed using the MIPS instrument team Data Analysis
Tool \citep{gordon05} for basic reduction (dark subtraction, flat
fielding/illumination correction). The known transient behaviors
associated with the Ge detectors were removed by time filtering the
data in the 70 \um default-scale mode, and by subtracting the
off-source chopped background observations for data in the 70 \um
fine-scale mode. The processed data were then combined using the World
Coordinate System (WCS) information to produce final mosaics with
pixels half the size of the physical pixel scale.

The photometry for each target was extracted using aperture photometry
with multiple aperture settings as well as point-spread-function (PSF)
fitting at both bands. By averaging the photometry using different
methods (including multiple apertures), the estimated error provides a
better estimate of the effects of image quality, background noise, and nearby
contaminating sources. For aperture photometry at 24 \mm, we first
determined the centroid of each target by fitting a 2-D Gaussian core,
then computed the averaged integrated flux within a large aperture (a
radius of 14\farcs94 with sky annulus between 29\farcs88 and
42\farcs33), and a small aperture (a radius of 6\farcs23 with sky
annulus between 19\farcs92 and 29\farcs88). Because the PSF at 70 \um
is not well-sampled and the {\it Spitzer} pointing is good to within
1\arcsec, for aperture photometry at 70 \um default scale we used the
WCS information to place the aperture center and measured total flux
for two different aperture settings: a large aperture (a radius of
29\farcs55 with sky annulus between 39\farcs40 and 68\farcs95) and a
small aperture (a radius of 15\farcs96 with sky annulus between
18\farcs03 and 39\farcs01). At 70 \um fine-scale mode, we determined
the centroid the same way as for the 24 \um data, but only used a
single aperture (a radius of 16\arcsec with sky annulus between
19\arcsec and 39\arcsec) due to the limited field of view. We
determined the aperture correction based on the theoretical STinyTim
PSFs \citep{krist02} that were smoothed to match the observed PSFs.
An aperture correction of 1.143 (1.298) for the large aperture and
1.699 (1.972) for the small aperture was applied to the aperture
photometry at 24 \um (70 \um default-scale), respectively. A value of
1.933 for the 70 \um fine-scale mode was used for the aperture
correction.

We used, {\it StarFinder} \citep{diolaiti00} to extract point-source
photometry via PSF fitting. The PSF used at 24 \um was constructed
based on isolated calibration stars while at 70 \um we used the
smoothed STinyTim PSFs. For bright targets without other nearby
contaminating sources, the results from aperture photometry and PSF
fitting agree within 1\% at 24 \um and 2-5\% at 70 \mm. Some stars in
our sample are located in high cirrus regions; however, the background
variation at 70 \um is generally less than 10\% across the whole field
(5\arcmin~by 2.5\arcmin). For the targets that have higher cirrus
variations, we made sure that the extracted source position agrees
within $\lesssim$1\arcsec (a general {\it Spitzer} pointing error)
between 24 and 70 \mm. The final measured flux and uncertainty were
determined by averaging the different photometry methods. Conversion
factors of 1.05$\times10^{-3}$, 16.5, 61.6 mJy/arcsec$^2$ were used to transfer
measured instrumental units to physical units (mJy) for 24 \mm, 70 \um
default-, and fine-scale modes, respectively. The absolute flux
calibration errors are less than 5\% at 24 \um and 10\% at 70 \um (Engelbracht
et al. 2006, in preparation; Gordon et al. 2006, in preparation). The
errors presented here are the intrinsic noise from images, and do not
include the calibration errors.

At 24 \mm, the majority of the stars in the sample have high ($\gg$20)
signal-to-noise detections, except those that have contamination from
nearby back/foreground sources or nebulosity. For our 70 \um
observations, all stars observed in the GTO Astar program used the
default-scale photometry AOT.  The integration times for these
observations were planned to provide at least 1-$\sigma$ detection of
the photosphere (including detector and background noises), but due to
various circumstances of observations and general overestimations of
the {\it Spitzer} sensitivity, many of these sources were not detected
at 70 \mm.  Observations from the other GTO programs (Fab4 and
Dirty12) targeted known {\it IRAS} debris disk candidates; therefore,
the integration time was designed for high signal-to-noise
detections. Finally, observations from the calibration program were
aimed to detect the predicted photospheres at least at the 3-$\sigma$
level.

In all cases, a threshold of 3-$\sigma$ was set for source
detection. If the signal-to-noise ratio is less than 3, we use the
3-$\sigma$ flux as an upper limit. We show in Table \ref{tab_sample}
the final measurements ($F_{m,24}$ and $F_{m,70}$ for 24 and 70 \mm,
respectively), uncertainties ($\sigma_{24}$ and $\sigma_{70}$) and
signal-to-noise ratios (SN$_{24}$ and SN$_{70}$) for the entire
sample.

\begin{figure}
\figurenum{1}
\label{agedistr} 
\plotone{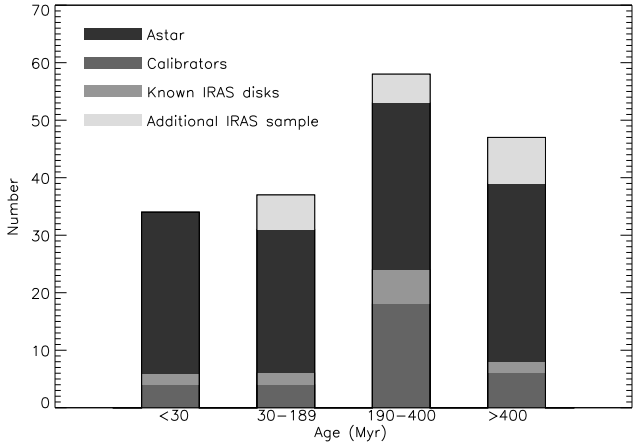}
\caption{The age distribution of the complete sample.}
\end{figure}

\subsection{Supplemental {\it IRAS} Sample}
\label{additional_excess}

The stars observed with {\it Spitzer} were selected on the bases of
(1) observability (background brightness and stellar magnitude) and
(2) availability of accurate age estimates (membership in cluster or
moving groups, etc.). There are a number of stars (47) that are
logical members of our sample that have not been observed with {\it
Spitzer} primarily because of the lack of accurate age
estimates. Generally, the {\it Spitzer} programs emphasized young
stars with ages from membership in moving groups or clusters, while these
unobserved stars tend to be relatively old according to ages estimated
from their placement on the HRD \citep{rieke05}. We added them
to this study using {\it IRAS} data to avoid any age bias. 
To do so, we proceeded as follows.

Given the large {\it IRAS} beams, we were concerned that noise
associated with the background might produce false excesses.  The
background (IR cirrus) can be estimated either directly from models of
the infrared sky (e.g., from the {\it Spitzer} Science Center SPOT
user tool) or from the atomic hydrogen column, n$_{\rm H}$ (e.g.,
Chandra Colden: Galactic Neutral Hydrogen Density
Calculator\footnote{http://cxc.harvard.edu/toolkit/colden.jsp}). We
used both methods to guard against stars in high cirrus regions; in
general they gave consistent results.  To determine a threshold for
false excesses, we examined the $K_s-[25]$ colors of the full sample
of stars as a function of the two background estimators. We concluded
that stars with n$_{\rm H} <$ 10 cm$^{-2}$ and IR background less than
1.3 MJy/sr at 60 \um are unlikely to suffer from cirrus confusion, and
we trimmed the sample of 47 stars to 30 that meet both criteria. We
next determined which of these stars have useful {\it IRAS}
measurements at 60 \mm. We used the Faint Source Catalog (FSC) to
obtain a 60 \um flux density measurement and an error in the
measurement for each star. Most of the stars are not detected, but the
errors allow setting upper limits to their fluxes. We trimmed the
sample to the 19 stars where there are either detected excesses, or we
could set 2-$\sigma$ upper limits to the excesses at less than 5 times
the photospheric flux density (Adopting 2-$\sigma$ limits allows
meaningful limits in the same range as the {\it Spitzer} measurements,
with a chance of one or two excesses slightly exceeding the actual
limits for those 19 stars). Because the two cuts are based on sky
properties, and then on {\it IRAS} signal to noise, the remaining
stars should be representative of the original sample of twice as
many. The properties of these 19 stars are listed in Table 2. The age
distribution of the complete sample is shown in Figure \ref{agedistr}.

\section{Photospheric Prediction and Excess Determination} 

\subsection{Excess Identifications and Statistics}

To determine the excess emission from the debris, the stellar
contribution has to be subtracted from the measurements. We determined
the stellar contribution at each band using the best-fit synthetic
Kurucz model \citep{castelli03} by fitting all available optical to
near infrared photometry (Johnson $UBVRIJHK$ photometry, Str\"{o}mgen
$uvby$ photometry, Hipparcos Tyco $BV$ photometry, 2MASS $JHK_s$
photometry) based on a $\chi^2$ goodness of fit test. 
For stars within
50 pc of the Sun, no correction for interstellar extinction was
applied. For stars with distances larger than 50 pc, we estimated the
extinction ($A_V$) based on the $B-V$ color and spectral type, and
then applied a reddening correction using the extinction curve from
\citet{cardelli89}. The predicted flux was then computed using the
best-fit Kurucz model spectrum at the 24 and 70 \um  weighted average
wavelengths (23.68 and 71.42 \mm, respectively).

\begin{figure}
\figurenum{2}
\label{hist_frac24} 
\plotone{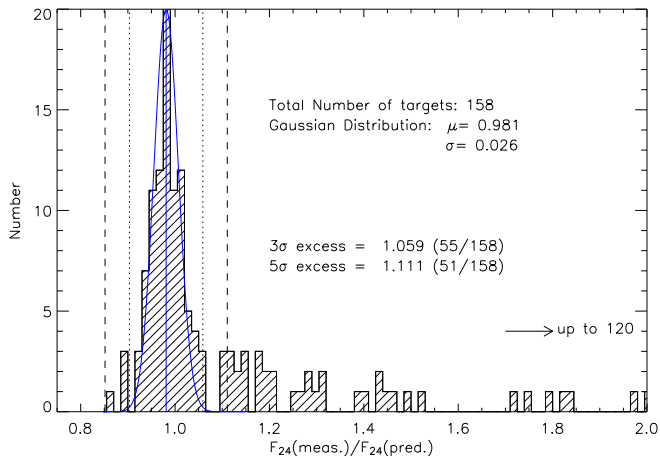}
\caption{Distribution of the 24 \um fluxes relative to the expected
  photospheric values from Kurucz model fittings. A Gaussian
  distribution with a dispersion of 0.026 is shown for
  comparison. Boundaries for $\pm$3$\sigma$ (dotted lines) and
  $\pm$5$\sigma$ (dashed lines) are marked.}
\end{figure}

To determine whether a star possesses a significant infrared excess,
we first need to evaluate how good our photospheric predictions and
measured photometry are. Figure \ref{hist_frac24} shows a histogram of
the 24 \um fluxes ratioed to the expected photospheric values. A
Gaussian distribution with a dispersion of 0.026 is shown for
comparison, indicating our predicted photospheres and measured
photometry are as good as 2.6\% at the 1-$\sigma$ level. 
A few outliers of low flux ratio values are probably due to the
effects of latent images (Engelbracht et al. 2006, in preparation).
Note that the center of the Gaussian is at 0.981, suggesting a
systematic offset of $\sim$2\%. Based on this we define a 3-$\sigma$
excess as a ratio greater than 1.06 or a 5-$\sigma$ excess as a ratio
greater than 1.11. There are four stars (HD 2266, HD 27045, HD 106591,
and HD 74956) that have ratios between 3- and 5-$\sigma$, and three of
them have confirmed 70 \um excesses (see below).
Therefore, we adopt the ratio of 1.06 (3-$\sigma$ excess) as our
threshold to identify infrared excesses at 24 \mm.

\begin{figure}
\figurenum{3}
\label{hist_frac70} 
\plotone{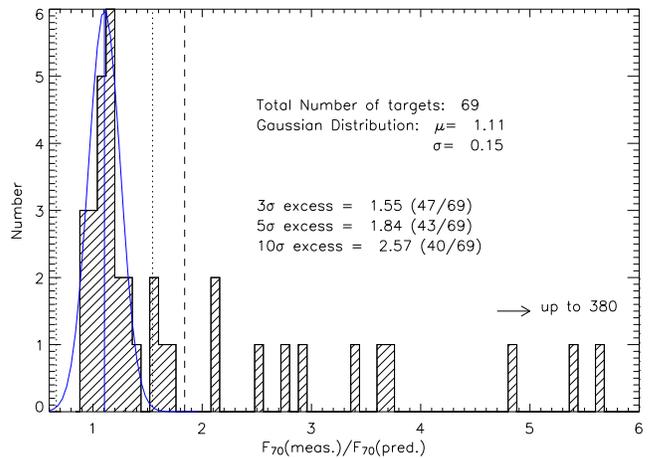}
\caption{Distribution of the 70 \um fluxes relative to the expected
  photospheric values from Kurucz model fittings. A Gaussian
  distribution with a dispersion of 0.15 is shown for
  comparison. Boundaries for $\pm$3$\sigma$ (dotted lines) and $\pm$5$\sigma$ (dashed
  lines) are marked.}
\end{figure}

A total of 137 stars have MIPS 70 \um observations, but only 69 stars
have positive detections (signal-to-noise larger than 3).  Figure
\ref{hist_frac70} shows a histogram of the 70 \um fluxes relative to
the expected photospheric values. A Gaussian distribution with a
dispersion of 0.15 and centered at 1.11 is shown for comparison. The
majority of the stars (40 out of 69) show large excesses
($>$10-$\sigma$).  We define a 3-$\sigma$ excess as a ratio between
measured and predicted fluxes greater than 1.55, and a 5-$\sigma$ excess
as a ratio greater than 1.84 at 70 \mm. Four stars have ratios that
fall between 3- and 5-$\sigma$ excesses; and three of them (HD 19356,
HD 106591, HD 115892) have 24 \um excess above 3-$\sigma$ (the
exception is HD 4150).  The 3-$\sigma$ threshold at 70 \um is a
consistent cutoff with the 3-$\sigma$ threshold at 24 \um because none
of the stars (a total of 22) that have ratios less than our 3-$\sigma$
threshold at 70 \um has a 24 \um excess (more than 3-$\sigma$).

Using these criteria (3-$\sigma$ excess as a flux ratio higher than
1.06 at 24 \um and 1.55 at 70 \mm), each of the stars is then
classified as ``YES'', ``NO'' or ``UPL'' in Table \ref{tab_sample}
corresponding to having an infrared excess, being detected but with no
excess above the 3-$\sigma$ confidence level, or only an upper
limit. In addition, we also used the significance ($\mathcal{X}$) of a
detected excess, defined as (measured$-$predicted)/uncertainty, for an
internal check. At 70 \mm, the excess stars identified by our flux
ratio criterion all have $\mathcal{X}_{70} \ge$3. At 24 \mm, two stars
(HD 93738 and HD 137919) have the excess significance
$\mathcal{X}_{24}$ less than 3 while the rest all have
$\mathcal{X}_{24} \ge$3. Unfortunately, the 70 \um measurements of
these two stars are both upper limits, and cannot verify the infrared
excess nature of these two stars. We, therefore, disregard HD 93738
and HD 137919 as having 24 \um excesses. At the 3-$\sigma$ confidence
level, the 24 \um excess rate is 32$\pm$5\% out of 155 stars (three
stars are not debris disks in nature, see Sec \ref{gas_disks}).  If we
assumed that all the non-detected stars do not have 70 \um excesses,
then the 70 \um excess rate is 33$\pm$5\% (out of 134). This is a
lower limit since in some cases the measured noise is much higher than
the predicted photospheric flux.  Therefore the excess rate at 70 \um
could be as high as 67$\pm$10\% (out of 66).  Among the 44 stars that
have excesses at 70 \um above the 3-$\sigma$ confidence level, eight
of them show no (less than 3$\sigma$) 24 \um excess. These eight
only-70\um excesses are unlikely to be false detections because the
significance of the detected excesses are all greater than 4 at 70 \um
(See Table 3, group II).

The fact that the excess rate at 24 \um and the lower-bound excess
rate at 70 \um are similar indicates
that MIPS 24 \um photometry is a very powerful and reliable tool to
study infrared excesses around early-type stars.  Combining both 24
and 70 \um results, the infrared excess detection rate is at least
37$\pm5$\% (58 out of 157) around A-type stars. Even using the same
criteria (R$_{24}> \sim1.2$ and R$_{70}> \sim1.5$) for identifying
excesses for the field (old) FGK stars (12$\pm$4\%, \citealt{bryden06})
and M-type stars ($\sim$0\%, \citealt{gautier06}), the
excess detection rate around A-type stars ($\sim$30\%, using the same
thresholds) is considerably higher.

\subsection{Decay Times}
\label{decay_time}

\begin{figure}
\figurenum{4}
\label{age_frac24} 
\plotone{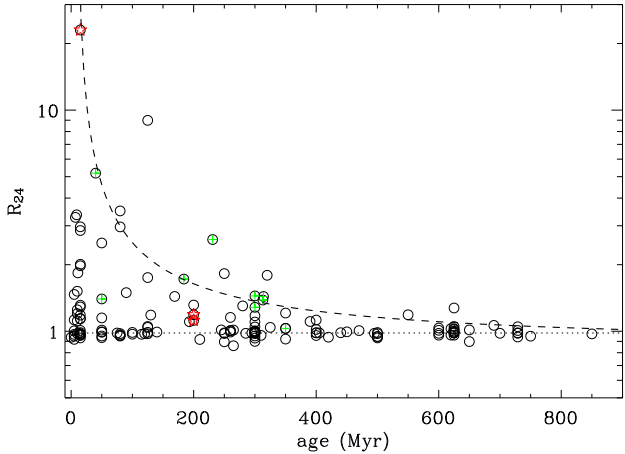}
\caption{24 \um excess vs.~age. Excess emission is indicated as the ratio
  of the observed flux density to that expected from the stellar
  photospheric value. $\beta$ Pic, Vega and Fomalhaut are additionally
  marked as star-shaped symbols while stars from the {\it IRAS}
  discovered debris disks are shown as plus signs. A decay curve of
  $t_o/t$ is indicated as the dashed line with $t_o \sim$150 Myr.}
\end{figure}

\begin{figure}
\figurenum{5}
\label{age_frac70} 
\plotone{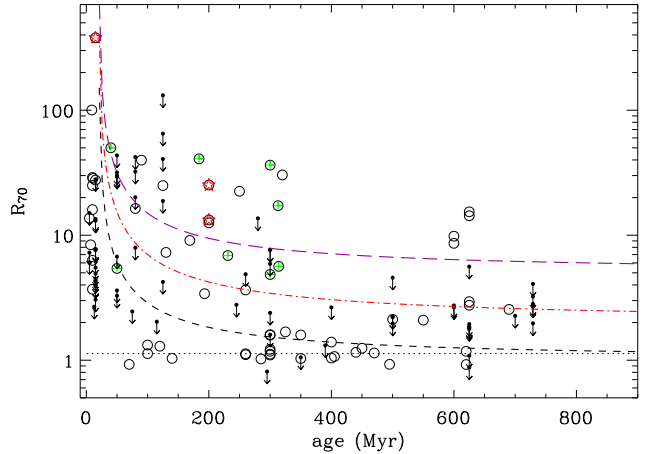}
\caption{Similar to Figure \ref{age_frac24} but for 70 \um excess
  vs. age. The upper limits are shown in small dots with downward
  arrows. Three decay curves of $t_o/t$ are indicated as the dashed
  line for $t_o \sim$ 150 Myr, the dot-dashed line for $t_o \sim$ 400
  Myr, and the long-dashed line for $t_o \sim$ 800 Myr.}
\end{figure}

The distributions of 24 and 70 \um excesses with stellar age are
illustrated in Figures \ref{age_frac24} and \ref{age_frac70}.
Twelve nearby debris disk stars discovered by {\it IRAS} are indicated
by star-shaped symbols (for $\beta$ Pic, Vega and Fomalhaut) and plus
signs (for HD 14055, HD 18978, HD 38678, HD 74956, HD 95418, HD
102647, HD 139006, HD 161868, HD 181296). As has been found in
\citet{rieke05}, the amount of excess emission at 24 \um shows a rapid
decline with stellar age ($\sim t_o/t$ and $t_o \sim$ 150 Myr), and a
large variety of excess amounts at any given age. The trend at 70 \um
also shows a large variety at any given age in Figure
\ref{age_frac70}. However, compared to the trend at 24 \mm, the 70 \um
excess trend has a much longer decay time with $t_o \ge$ 400 Myr.

We have carried out a simple analysis to show that the trends for
different decay times at 24 and 70 \um evident in Figures
\ref{age_frac24} and \ref{age_frac70} are statistically
significant. We divided the sample into two parts, based on the 24 \um
decay time scale of $\sim$150 Myr \citep{rieke05}.  Given the age
uncertainties, we put the division at 400 Myr.
For the stars with ages less than this value, we determined that the
proportion of stars with excess ratios $>$1.3 at 24 \um and $>$5 at 70
\um were virtually identical (that is, 28 out of 109 and 26 out of 95,
respectively). However, for the stars 400 Myr old or older, there were
0 stars out of 45 with excess ratios $>$1.3 at 24 \um, but 4 stars out
of 38 with excess ratios $>$5 at 70 \mm. We used the binomial theorem
to show that the probability of these two results coming from the same
parent distribution is only about 1\%.  That is, the 24 \um excesses
decay more rapidly than those at 70 \um at the 99\% confidence level.

\section{Dust properties around A Star Debris disks}

\subsection{Circumstellar Gas Disks } 
\label{gas_disks}
\begin{figure}
\figurenum{6}
\label{sedgasdisks} 
\plotone{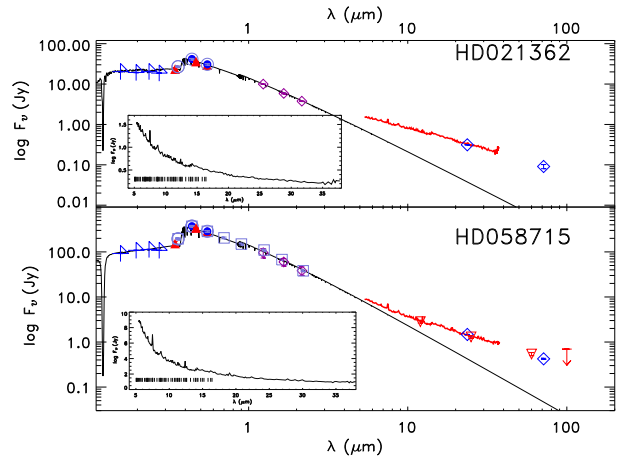}
\caption{Spectral Energy Distributions for Be stars HD 21362 and
  HD 58715. Data plotted are: TD1 UV fluxes (plus signs), {\it uvby}
  photometry (filled triangles), Johnson UBV (large open circles),
  Hipparcos photometry (small filled circles), 2MASS photometry (open
  diamonds), IRAS color corrected fluxes (downward triangles or downward
  arrows for upper limits), MIPS fluxes (open blue diamonds). The IRS
  spectra are shown in red lines between 5 and 38 \mm, as well as the
  small insets in a linear scale.   
  The wavelengths of the hydrogen lines between 5 and 17 \um
  are identified in the small inset. The prominent 
  line at 7.495 \um is the HI 18-8 line.}
\end{figure}

Two stars in our sample, HD 21362 and HD58715, are associated with the
Be phenomenon, i.e., where the strong stellar wind from a fast
rotating B-type star forms a circumstellar gas disk, showing hydrogen
emission lines in the optical and excess radiation relative to the
expected photospheric flux in the infrared. The infrared excess
emission is due to the free-free radiation from the ionized stellar
wind. The nature of these gas disks can be recognized by several
hydrogen lines seen in the IRS low resolution spectra (Figure
\ref{sedgasdisks}). The spectra were observed as part of the GTO
follow-up debris disk programs. Details of the observations and data
reduction will be discussed in an upcoming paper (Su et al. 2006, in
preparation).

These two stars as well as HD 58467 (a Herbig Ae/Be star) are not
debris disks in nature, and are disregarded in the following
discussion as well as the proceeding discussion on the infrared excess
rates.

\subsection{[24]$-$[70] Color temperature} 

In the following sections, we discuss the dust properties in terms of
the observed [24]$-$[70] color temperature and fractional dust
luminosity with stellar age.  There are 36 stars that have both 24 and
70 \um excesses, above 3-$\sigma$ confidence; hereafter we refer to
them as group I. The stars that only show 70 \um excess (a total of
8), we refer to as group II. Stars that have detections with S/N$>$3
at both 24 and 70 \um but have no excess above the 3$\sigma$ levels,
we refer to as group III (a total of 20). The infrared excess fluxes
($F_{ire,24}$ and $F_{ire,70}$ for 24 and 70 \mm, respectively),
significance of the excess fluxes ($\mathcal{X}_{24}$ and
$\mathcal{X}_{70}$), and the dust properties for stars from group I to
III are listed in Table 3. Stars that have no significant 24 \um
excesses and for which the 70 \um measurements are upper limits, we
refer to as group IV (a total of 55). Group V (a total of 13) are the
stars that have significant 24 \um excesses (above 3-$\sigma$
confidence levels) but their 70 \um measurements are upper limits.

\begin{figure}
\figurenum{7}
\label{bbt_dist} 
\plotone{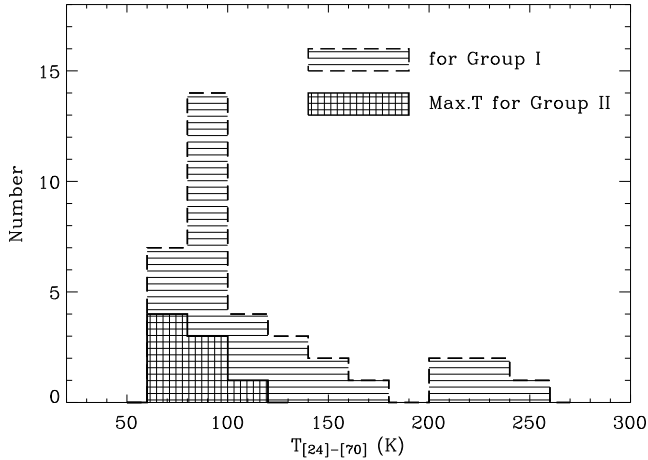}
\caption{The $[24-70]$ color temperature distribution of the
  disks. The stars that have both 24 and 70 \um excesses (group I)
  are displayed in a horizontal-lined pattern.
  The maximum color temperatures for stars in group II (only 70 \um
  excess) are shown in a mesh pattern.}
\end{figure}

One simple way to characterize the disk properties based on the 24 and
70 \um excess emission is a blackbody fit to determine the color
temperature (T$_{[24]-[70]}$) of the disk. It is straightforward to
determine the color temperature for group I stars by ratioing the flux
density at the two bands. Including the uncertainty in the excess
fluxes, the color temperature generally has an uncertainty of $\pm$5 K
for the group I stars. For the stars in group II (excess at 70 \um but
not at 24 \mm), a color temperature is computed by assuming that the
excess flux at 24 \um is three times the measured uncertainty, which
serves as a maximum temperature that the dust can have, consistent
with our photometric accuracy.  Figure \ref{bbt_dist} shows the color
temperature histogram for the group I and group II stars.

The majority of the debris disks have T$_{[24]-[70]}\sim$90 K, i.e.,
they are at a distance of $\sim$100 AU from the star if we assume that
the dust particles in these systems are blackbody-like. The Kuiper
belt in these A stars is expected to extend about 1.6 times farther
than the Sun's (45-55 AU) following a simple mass scaling
\citep{su05}.  Hence, most of the debris around these A stars is
Kuiper-Belt-like if it consists of large (radius $\ge$50 \mm)
blackbody radiators. However, based on the study of the Vega debris
disk \citep{su05}, non-blackbody-like small grains (radius $\le$ 10
\mm) can dominate the disk radiometric properties; therefore, the disk
size can extend a few times larger if the dust grains in the system
are small.

There are five stars that have [24]$-$[70] color temperatures larger
than 200 K: HD 19356, HD 23862, HD 38678, HD 75416, and HD
115892. Since both HD 21362 and HD 58715 (gas disks) also have
[24]$-$[70] color temperatures larger than 200 K, it is possible that
these 5 stars are gas disks as well. However, no gas lines were seen
in the mid-infrared spectra of HD 19356, HD 38678, and HD 115892
(\citealt{chen06}; Su et al.~2006, in preparation), which leaves only
HD 23862 and HD 75416 as possible gas disks.

\begin{figure}
\figurenum{8}
\label{age_bbt} 
\plotone{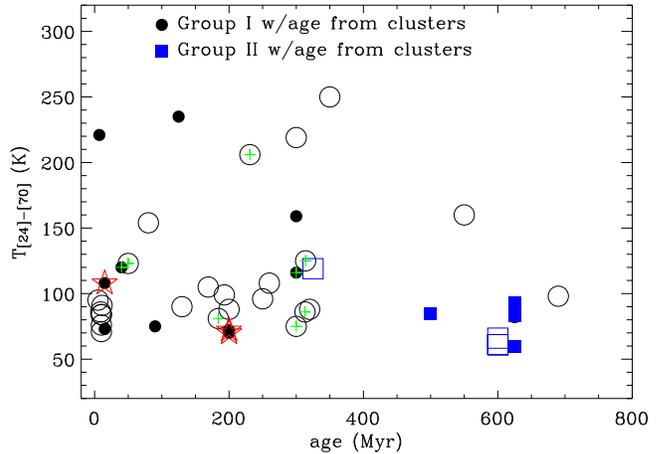}
\caption{Stellar age vs. observed [24]$-$[70] color
  temperature. Symbols used are circles for stars that have both 24
  and 70 \um excess (group I), squares for stars only show 70 \um
  excess (group II).  Filled and smaller-size symbols are for stars
  with age associated with clusters or moving groups. $\beta$ Pic,
  Vega and Fomalhaut are additionally marked as star-shape symbols
  while the stars from other IRAS discovered disks as plus signs.}
\end{figure}

Figure \ref{age_bbt} shows the distribution of the observed
[24]$-$[70] color temperature with stellar age. 
Slightly smaller, but filled symbols represent the stars with ages
determined from cluster associations.  The ages determined from
clusters or moving group association generally have errors less than
50\%, but the ages determined from the HRD could have errors up to a
factor of two. It appears that the color temperature of the disks has
a slightly broader distribution when stars are younger. For stars older than
400 Myr, most of the color temperatures of the disks appear in a
narrow range between 50-150 K. Assuming the dust we see is
blackbody-like (i.e., the location of the dust is directly related to
dust temperature and stellar properties), the decrease of the
[24]$-$[70] color suggests that the debris is located further away
from the stars in the older systems, consistent with the significant
difference in decay time scales between 24 and 70 \mm, found in Sec 3.2.

\subsection{Trends in Amounts of Excess Emission} 

Since the amounts of excess at both 24 and 70 \um decrease with
stellar age (shown in Sec \ref{decay_time}), it is 
important to have a large number of stars especially in the old age bin to
ensure good statistics. As stated in Sec \ref{additional_excess}, an
additional 19 A-type stars with {\it IRAS} observations at 25 and
60 \um were included to avoid an age bias. There are two candidates among
these stars to show possible excesses at 25 \mm: HD 56537, with a flux
ratio of 1.32, and HD 79469 with a flux ratio of 1.30. Both are in the
``marginal'' category, since the deviations from unity flux ratio (no
excess) are less than 3-$\sigma$. There is only one star with a
well-detected excess at 60 \mm, HD 39014 (see, e.g.,
\citealt{jura04}). HD 142105 has a possible excess at 60 \um (3.9
$\sigma$). For all the remaining stars, there are only upper limits at
60 \mm. To analyze all the stars in this sample and in our {\it
Spitzer} sample, in the following we will discuss 2-$\sigma$ upper
limits for {\it IRAS} measurements at a uniform level of 1.25 at 25
\um and 5 at 60 \mm.

\begin{figure}
\figurenum{9}
\label{ire_trends} 
\plotone{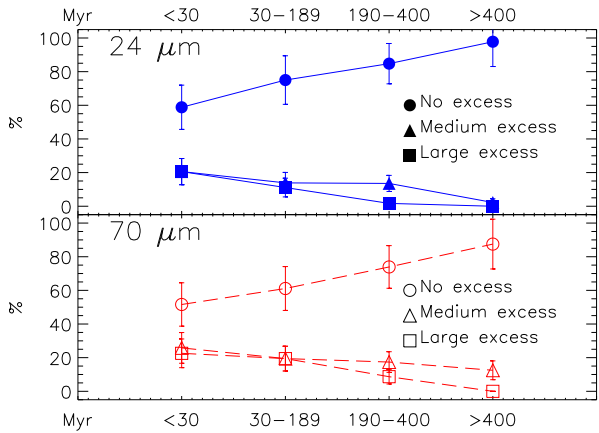}
\caption{Trends of excesses with age at both 24 and 70 \mm, based on
  Table \ref{excess_trend}. The 24 \um results are plotted as filled
  symbols with solid connecting lines while the 70 \um results are as
  open symbols with dashed connecting lines. ``No excess'' is shown in
  circles, ``Medium excess'' in triangles, and ``Large excess'' in
  squares. }
\end{figure}

These values correspond to a color temperature of $\sim$90 K, which is
typical of most of the debris systems (see Figure \ref{bbt_dist}). We
divide stars from the {\it Spitzer} sample into those with measured
excesses above these limits, or excess or upper limits below them
(i.e., R$_{24}>$1.25 and R$_{70}>5$ even with the 2-$\sigma$ upper
limits). Combining the {\it IRAS} and {\it Spitzer} samples, a total
of 174 stars were used at 24/25 \mm, and 153 stars for 60/70 \um to
study the trends in amounts of excess emission.

We further group the observations into bins in ages and amount of
excess for three subgroups: ``no'' means no excess; ``large'' means a
flux ratio greater than 2 for 24 \um and 20 for 70 \mm; ``medium''
means the flux ratio is intermediate. The age bins were set so that
the total number of stars in each age bin is roughly equal.  The
results are shown in Table 4 and Figure \ref{ire_trends}. The 24 \um
excess trend is consistent with the results of \citet{rieke05} that
$\sim$50-60\% of the stars that are younger than 30 Myr have no 24 \um
excess, rising to $\sim$85-95\% for the stars that are older than 190
Myr. The trend of no excess at 70 \um is similar to 24 \um within the
errors, but with a systematically lower fraction.

The trends of infrared excess (either medium or large) are, again,
similar, but the fractions at 70 \um are systematically higher than at
24 \mm, suggesting that the rate of 70 \um excesses is higher than at
24 \um and the persistence time of the 70 \um excess is longer; i.e.,
the debris primarily emitting at 70 \um remains in the system longer.

\begin{deluxetable}{ccccc}
\tablenum{4}
\tablecaption{Excess Amount v.s. Age\label{excess_trend}}
\tablewidth{0pt}
\tablehead{ 
\colhead{Myr}&\colhead{tot\#}&\colhead{No}&\colhead{Medium}&\colhead{Large}
}
\startdata 
for 24/25 \um         &     &             &          &       \\
   $  t \le$  30 & 34  &  20(58\%)   &  7(21\%) & 7(21\%)\\ 
 30$< t \le$ 189 & 36  &  27(75\%)   &  5(13\%) & 4(11\%)\\
190$< t \le$ 400 & 59  &  50(85\%)   &  8(14\%) & 1( 2\%)\\
400$< t   $      & 45  &  44(98\%)   &  1( 2\%) & 0( 0\%)\\
                 &     &             &          &        \\      
for 60/70 \um         &     &             &          &        \\
   $  t \le$  30 & 31  &  16(52\%)   &  8(25\%) & 7(23\%)\\
 30$< t \le$ 189 & 36  &  22(62\%)   &  7(19\%) & 7(19\%)\\
190$< t \le$ 400 & 46  &  34(74\%)   &  8(17\%) & 4( 9\%)\\
400$< t   $      & 40  &  35(88\%)   &  5(12\%) & 0( 0\%)\\
\enddata
\tablerefs{Definitions of ``No'', ``Medium Excess'' and ``Large
  Excess'' for flux ratio at 24 \um (R$_{24/25}$) and 70 \um
  (R$_{60/70}$) are: no -- R$_{24}<$ 1.25; large -- R$_{24}>$2;
   median -- 1.25$\le$R$_{24}\le$2; no -- R$_{70}<$ 5; 
   large -- R$_{70}>$20; median -- 5$\le$R$_{70}\le$20.}
\end{deluxetable}

\subsection{Fractional Dust Luminosity} 

The most frequently used quantity to measure the amount of dust in
these systems is the fractional luminosity, $f_d$: the ratio of the
total emission by dust to the stellar luminosity ($f_d = L_{IR}/L_{\ast}$). It
measures the fraction of the sky seen from the star that is covered by
dust, and therefore the fraction of the stellar radiation that will be
absorbed and re-emitted in the infrared \citep{dominik03}. The
fractional luminosity can be determined based on the observed
[24]$-$[70] color temperature and an assumption of blackbody emission
for the dust. However, the fractional luminosity estimated in this way
is somewhat over-estimated since the dust grains in the system
probably have a $\lambda^{-\beta}$ emissivity dependence where
$\beta=$1-2, and emit less efficiently at longer wavelengths than
blackbody emission. The overestimate could be large,
depending on details of the emissivity law. 
We discuss HD 225200 as an example in Sec 4.5; the
fractional luminosity is $\sim$8\% over-estimated between a blackbody
and modified blackbody emission. 

Large errors will result if these disks have substantial emission from
cold ($\sim$30 K) dust. In most cases, there are no observations to
constrain this possibility very well. However, examining the three
best-studied nearby A-type debris disks, Vega, Fomalhaut and $\beta$
Pic, we find the flux ratio between 70 and 850 \um is $\sim$140,
$\sim$92, and $\sim$125, respectively. Given the distinctly different
properties of these three disks (age, fractional luminosity, and disk
extent), an average flux ratio of $\sim$120 between 70 and 850 \um
should be representative for early-type debris disks. This average
flux ratio gives a dust temperature of $\sim$64 K assuming a $\beta$=1
emissivity law, and the resultant fractional luminosity is roughly
equal to or less than the one computed using the maximum color
temperature. Applying similar logic, an upper-limit fractional
luminosity is also estimated for each of the stars in group III using
a computed color temperature by assuming that the excess flux at each
band is three times the measured uncertainty.

\begin{figure}
\figurenum{10}
\label{f_dist} 
\plotone{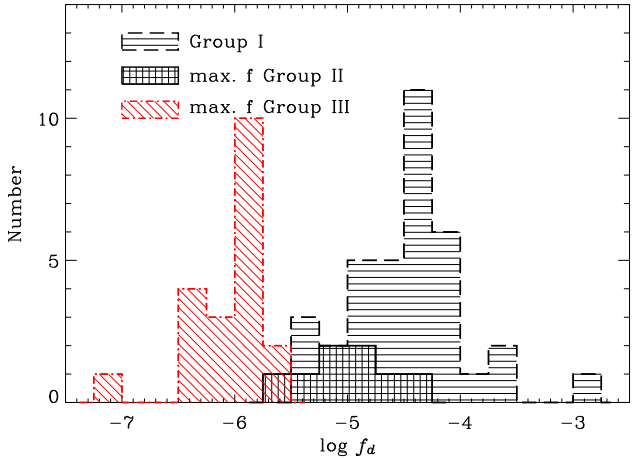}
\caption{The fractional infrared luminosity distribution of the debris 
  disks around A-type stars. The stars that have both 24 and 70 \um excesses (group I)
  are displayed in horizontal-lined pattern while the stars in group II (70
  \um excess only) are in a mesh pattern. The group III distribution
  shows that our detection limits are strong enough that there should be no
  significant bias in the group I/II comparison. }
\end{figure}

\begin{figure}
\figurenum{11}
\label{f_dist_agecut} 
\plotone{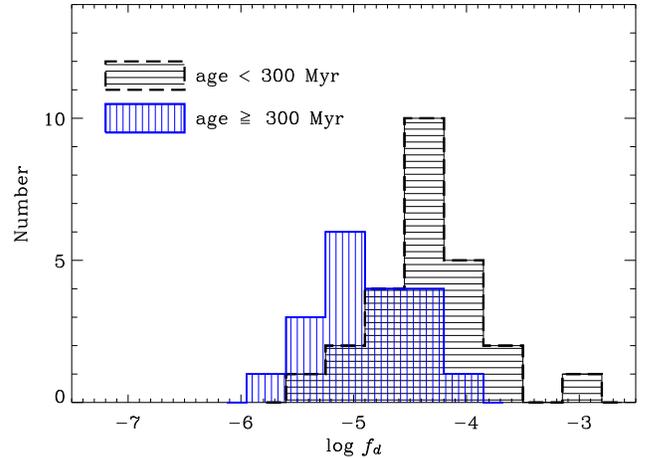}
\caption{Same as Figure \ref{f_dist} but with two age bins (older or
  younger than 300 Myr) among the group I and II stars.}
\end{figure}

A histogram of fractional luminosity is shown in Figure
\ref{f_dist}. The majority of the stars have a fractional infrared
luminosity $\sim5\times10^{-5}$. Stars that only show 70 \um excess
(group II) generally have lower fractional infrared luminosity. The
maximum fractional luminosities for group III stars are much lower
than the group I and II stars, consistent with their non-detectable
infrared excess. To see the general trend of the fractional luminosity
with age, we divide the group I and II stars into two age bins: older
or younger than 300 Myr. This division is roughly at half the main
sequence lifetime of an A0~V star (the histogram is not sensitive to
this age; it is similar if 
we make the cut at an age of 400 Myr). The fractional luminosity
distribution for these two age groups is shown in Figure
\ref{f_dist_agecut}; older stars tend to have lower fractional
luminosity than younger ones, which confirms the results from Sec
\ref{decay_time} and Figure \ref{ire_trends}.

\begin{figure}
\figurenum{12}
\label{age_f} 
\plotone{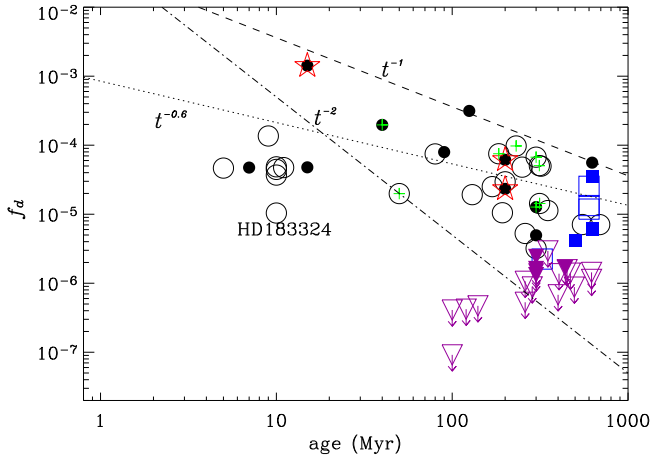}
\caption{Stellar age vs.~fractional luminosity. Symbols used are the
  same as Figure \ref{age_bbt}. Additional data points from group III
  stars (maximum detectable fractional luminosity) are shown as upside-down triangles with arrows pointing
  downwad, indicating our detection limits. 
The 1/$t$ relation is shown as a
dashed line while 1/$t^2$ as a dot-dashed line, with arbitrary
normalization. A 1/$t^{0.6}$ relation was obtained by fitting
all the confirmed excess stars (group I and II), shown as a dotted
line.}
\end{figure}

The distribution of the fractional luminosity with stellar age is
shown in Figure \ref{age_f}. Several characteristics are found in this
$f_d$ vs.~age diagram:

\begin{enumerate}
  \item The data are consistent with a general 1/$t$ relation in the
    $f_d$ vs.~age diagram but with at least two orders of magnitude
    variation in the amounts of $f_d$.

  \item An upper envelope of 1/$t$ is seen in Figure \ref{age_f}. We
    do not detect any stars that are older than 100 Myr and have $f_d$
    values greater than 10$^{-3}$ in this early-type sample.

  \item All the only-70\um excess stars (group II, squares in Figure
    \ref{age_f}) are old. No stars that are younger than
    100 Myr have only-70 \um excess. 
    This trend could result from sample
    selection bias, because most of the young stars are at larger
    distances, and only 70 \um upper limits were obtained. Among the 53  
    group IV stars (no 24 \um excess and 70 \um is an upper limit), 
    6 are younger than 100 Myr, have 3-$\sigma$ upper
    limit ratios larger than 5, and most importantly their backgrounds
    are clean based on the 24 \um images. These 6 stars are potential 
    only-70 \um excess young stars. Future deeper 70 \um observations
    can help to better identify their natures. 
    
  \item Stars that have no detectable excess (group III, upside down
    triangles with downward arrows in Figure \ref{age_f}) have $f_d$ as
    low as $\sim$10$^{-7}$ (similar to the value in our Solar System). 

\end{enumerate}

A constant upper envelope of $f_d \approx$10$^{-3}$ is suggested by
\citet{decin03} based on a re-analysis of the data obtained with
ISOPHOT and revised age estimates. This upper constant cutoff is not
seen in these new {\it Spitzer} observations (item 2). Although Decin et
al.~combine all the available spectral types on the same plot, it is
possible that the maximum fractional luminosity is different for
different spectral types.

\citet{dominik03} suggest that the total disk mass in a system and
the location and sizes of the parent bodies are the three major
parameters in determining the place in the $f_d$ vs.~age diagram. They
conclude that a 1/$t$ relation indicates the dust removal process is
dominated by collisions, and that Pointing-Robertson (P-R) drag would
yield a 1/$t^2$ relation. Collisions which result in small grains that
are blown out of the system via radiation pressure are the dominant
mechanism in removing dust in the bright debris disk systems observed
by {\it IRAS} and {\it ISO}. This is true for these early-type stars
since blowout occurs for grains $\lesssim$10 \mm; however, 
radiation pressure from lower
luminosity stars (late K and M dwarfs) may not be adequate to remove
grains at all, and stellar wind drag is the dominant mechanism to
remove grains in the young late K to M dwarfs \citep{plavchan05} . 
Hence, for the A - G systems, P-R drag only plays an important role
when the density of the debris is as low as in our solar system
\citep{dominik03,wyatt05}. Therefore, most (if not all) of the stars
in the $f_d$ vs.~age diagram should follow a 1/$t$ trend. This is
consistent with what we see in these new {\it Spitzer} data.

Assuming that the group IV stars have
no 70 \um excess as well, the true lower envelope in the $f_d$ vs.~age
diagram is then very low. The total surface area occupied by dust may
be greatly enhanced by collisions, and then follow a 1/$t$ steady state
collisional cascade. If the only-70 \um excesses are, indeed, 
only found associated with older stars, it means that the clearing
process (collisions) is an inside-out process. The collision frequency
is higher closer to the stars because of larger relative velocities;
therefore, the dust will be ground down to finer debris (subject to
radiation blowout) faster. This is consistent with what
\citet{wyatt05} has suggested. 

Alternatively, the ``clearing'' of the inner region could also arise
from the decline of inward grain transport, rather than the removal of the
initial inner disk grain population. A secular decline in the
collision rate in the outer disk might account for a drop in inward
grain transport, and thus a drop in 24 \um emission with time. A
crucial question for this model is whether the particles generated at
the 70 \um emission zone (50-200 AU for A stars) have enough time to
drift inward before they get destroyed by collisions. Assuming an A-type
star of 2.5 M$_{\sun}$ and 60 L$_{\sun}$ with a debris disk of $f_d
\sim5\times10^{-5}$, the P-R lifetime ($\sim10^{7-8}$ yr) is roughly
100 times the collisional lifetime ($\sim10^{5-6}$ yr) for grains with
radii $\sim$100 \um and density of 2.5 g/cm$^3$ at distances of
50-200 AU from the star. 
Therefore, the population of the particles that drift inward
from the 70 \um zone to the 24 \um emission zone is likely to be
small. Furthermore, we do not expect the color temperature to evolve 
with time if the 24 \um emission comes from material that is spiraling
in from the colder, outer disk via P-R drag. The amount of the dust we
see in these systems suggests the dominant particle removal mechanism
is through collisions. P-R drag may only be important 
for older systems because it does not explain
the dust temperature trend seen in the systems from 5 to 850 Myr of
age.

\subsection{Debris disk Model: Vega-like grains or KBO-like grains}

As has been thoroughly discussed in the literature, debris disk
modeling based on the broadband SED alone is degenerate, and hence not
conclusive in constraining the debris distribution or dust
properties. Based on the resolved disk surface brightness profiles
(100-800 AU), \citet{su05} have shown that the majority of the dust
particles in the disk are small grains in the proto-type Vega debris
disk. In comparison, the Vega SED has been modeled using large 30-200
\um grains in a ring-like (80-120 AU) disk \citep{dent00}. Resolving
the disk extent is important.  Unfortunately most of the debris disks
we discuss here are not resolved with {\it Spitzer}'s beams.
Without further constraints, there are many degeneracies using two
data points (24 and 70 \mm) to constrain a (at least) 6-parameter disk
model (surface density power-law index $p$, grain size power-law index
$q$, grain size limits ($a_{min}$ to $a_{max}$), disk extent ($R_{in}$
to $R_{out}$)).
 
\begin{figure}[t]
\figurenum{13}
\label{sedexample} 
\plotone{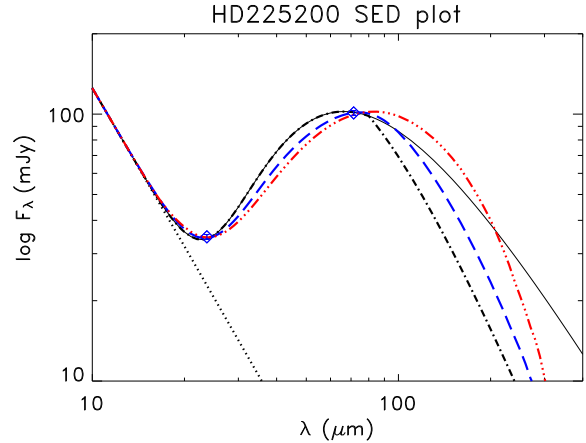} 
\caption{Model SEDs for HD 225200. The black dotted line represents the
  stellar photosphere. 
The black solid line is the SED
  assuming blackbody emission at  $T_d=75$ K while the dash-dotted black
  line is assuming a modified blackbody emission
  $(\frac{\lambda}{\lambda_o})^{-1} B_{\lambda}(75 K)$ with $\lambda_o
  =$80 \mm. The blue long dashed line is the best-fit SED using KBO-like
  grains while the red dash-triple-dotted line is the best-fit SED using Vega-like
  grains. Both models go through the 24 and 70 \um points, but have
  noticeably different shapes between these points. 
} 
\end{figure}

As an illustration, we fit the observed SED of HD 225200 with two
different pre-determined models: Vega-like or KBO-like, using
astronomical silicate grains. In the Vega-like model, we restrict the
grains to be ``small'' (i.e., $a_{min}\sim 1$ \mm, $a_{max}<$50 \mm,
$q=-2.5$), and to be driven outward by radiation pressure, therefore,
$p=1$. In the KBO-like model, we restrict the grains to be ``large''
(i.e., $a_{min}\sim 10$ \mm, $a_{max}=$100 \mm, $q=-3.5$), and to
spiral inward due to the P-R drag, therefore, $p=0$. The remaining two
parameters to fit are $R_{in}$ and $R_{out}$.

We searched large regions of parameter space, computing a $\chi^2$
statistic at each point. At 90\% confidence, we find that
$R_{in}=35\pm 15$ AU, $R_{out}=120\pm 117$ AU with dust mass of
$8.5\pm4.3 \times10^{-3} M_{\earth}$ for the KBO-like model, and
$R_{in}=100\pm 55$ AU, $R_{out}=900\pm 450$ AU with dust mass of
$21.40\pm5.2 \times10^{-3} M_{\earth}$ for the Vega-like model. Figure
\ref{sedexample} shows the resultant SEDs. The outer radius in the
Vega-like model ($\sim$900 AU, $\sim$7\arcsec~given a distance of 129
pc) is theoretically resolvable with MIPS at 24 \mm. The fact that our
shallow 24 \um observation indicates HD 225200 to be an unresolved
point source indicates either that deeper imaging might be required to
detect the disk outer boundary, or that Vega-like grains are not
likely present in the system. Furthermore, the two models yield
different amounts of emission at wavelengths greater than $\sim$100
\mm. Without resolved images, data points at far-infrared and
sub-millimeter wavelegnths can provide further constraints on the disk
outer boundary. Follow-up IRS observations of these A-type debris
disks can also put additional constraints on the disk model (inner
radius and grain sizes), and will be addressed in another paper.

\section{Summary and Conclusion} 

About 160 A-type main-sequence stars with ages ranging from 5 to 850
Myr were measured at 24 and/or 70 \um using the Multiband Imaging
Photometer for {\it Spitzer}. With {\it Spitzer's} unprecedented
sensitivity, we are able to identify infrared excesses of $\sim$6\% at
24 \um and $\sim$55\% at 70 \um above the photospheres at the
3-$\sigma$ levels. At this 3-$\sigma$ confidence level, the infrared
excess rate is 32$\pm$5\% and $\ge$33$\pm$5\% for 24 and 70 \mm,
respectively. The excess detection rate around these early-type stars
is significantly higher than the one found in the old FGK
\citep{bryden06} and M-type stars \citep{gautier06}. However, recent
studies by \citet{gorlova06} and \citet{siegler06} show that the
excess rate for young FGK stars is also higher than for old FGK stars;
most importantly, the evolution in the trend of excess rate vs.~age
looks similar between A stars and FGK stars (only the absolute
fraction is different) in their studies.  It is difficult to determine
whether there is an intrinsic difference in the infrared excess rates
between A-type and FGK stars because equivalent levels of fractional
luminosity become increasingly hard to detect as the star becomes
cooler. The incidence of debris disks is actually the convolution of
how many stars actually possess debris disks with given properties,
and how detectable the disks are with {\it Spitzer}, and this
convolution is a strong function of spectral type. Thus debris-disk
excess emission is not a steep function of stellar type. Since the
detectability of a given level of fractional excess decreases with
decreasing stellar temperature, the incidence of excesses is
consistent with a similar debris disk driven fractional excesses
between A and K stars. Nevertheless, it appears that age is the most
important factor in determining the detectability of infrared excess
among A- to K-type main-sequence stars, not stellar mass.

The amount of excess emission decreases with stellar age and follows a
simple $t_o/t$ relationship in general with $t_o \sim$150 Myr for
excesses at 24 \um but $t_o \ge$400 Myr at 70 \mm; that is, 24 \um
excesses decay more rapidly than those at 70 \mm. In addition, at any
given age there exists a large variety in the amount of 24 and 70 \um
excess emission. The observed [24]$-$[70] color temperatures for a
total of 44 debris disks with 24 and/or 70 \um excesses range from
$\sim$60 to $\sim$250 K, but the majority of the disks have a typical
temperature of $\sim$ 90 K. Furthermore, older stars (age $>$ 300 Myr)
tend to have colder observed [24]$-$[70] color temperatures than young
ones. Assuming the dust we detect is blackbody-like, the decline of
the [24]$-$[70] color implies the debris is located further away from
the stars in the older systems.

The fractional luminosity for these 44 early-type debris disks ranges
from $\sim$10$^{-3}$ to $\sim$10$^{-6}$. The trend between the
fractional luminosity and stellar age follows a general 1/$t$
relationship; older stars tend to have lower fractional luminosity
than younger ones. An upper envelope of 1/$t$ is seen in the
fractional luminosity vs.~age diagram, suggesting that collisions
followed by radiation pressure blowing out small grains are the
dominant process to remove grains in these systems. The decreases of
observed [24]$-$[70] color temperatures and the fractional luminosity
suggest that the debris disk clearing is an inside-out process.

\acknowledgments

Based on observations with NASA {\it Spitzer} Space Telescope, which
is operated by the Jet Propulsion Laboratory, California Institute of
Technology under NASA contract 1407.  Support for this work was
provided by NASA through Contract Number 1255094 issued by
JPL/Caltech.  This research made use of the SIMBAD database, operated
at CDS, Strasbourg, France. This publication makes use of data
products from the Two Micron All Sky Survey, which is a joint project
of the University of Massachusetts and the Infrared Processing and
Analysis Center/California Institute of Technology, funded by the
National Aeronautics and Space Administration and the National Science
Foundation.

\begin{turnpage}

\begin{deluxetable}{lcrrcrlrrrrcrrrrc}
\tabletypesize{\scriptsize}
\tablewidth{0pt} 
\setlength{\tabcolsep}{0.02in}
\tablenum{1}
\tablecaption{Sample of Stars and MIPS Measurements\label{tab_sample}}
\tablecolumns{19}
\tablehead{
\colhead{Name} & \colhead{AOR Key} & \colhead{SpType} &  \colhead{distance} & \colhead{Cluster/} & \colhead{Age}  &\colhead{Age ref.} &\colhead{F$_{m,24}$}&\colhead{$\sigma_{24}$}&\colhead{SN$_{24}$}&\colhead{R$_{24}$}&\colhead{IRE$_{24}$}& \colhead{F$_{m,70}$}&\colhead{$\sigma_{70}$}&\colhead{SN$_{70}$}&\colhead{R$_{70}$}&\colhead{IRE$_{70}$}
\\
\colhead{} & \colhead{} & \colhead{} &  \colhead{pc} & \colhead{Moving Group} & \colhead{Myr}  &\colhead{} &\colhead{mJy}&\colhead{mJy}&\colhead{}&\colhead{}&\colhead{}& \colhead{mJy}&\colhead{mJy}&\colhead{}&\colhead{}&\colhead{}
}
\startdata
HD000319  &   3972864 &      A1V &   80.3&       \nodata &   600& 3          &   43.69 &     0.78 &    56.01 &    0.979 &       NO &  $<$13.6 &     4.53  &       3 &    2.742 &      UPL  \\ 
HD001160  &  10090496 &      A0V &  136.6&       \nodata &     0$^{\flat}$& 22         &   10.56 &      0.2 &     52.8 &    0.939 &       NO &   \nodata&   \nodata &  \nodata&   \nodata&   \nodata \\ 
HD002262  &   3692544 &      A7V &   23.5&       \nodata &   690& 22         & 304.61  &   4.05   &  75.21   &  1.063   &    YES   &   78.7   &   3.74    & 21.04   &  2.548   &    YES     \\   
          &   6036736 & & & & & & & & & & & & & & & \\
          &   3692544 & & & & & & & & & & & & & & & \\
          &   6036736 & & & & & & & & & & & & & & & \\
HD002811  &   9940224 &      A3V &  205.8&       \nodata &   750& 22         &   10.56 &     0.07 &   150.86 &     0.95 &       NO &   \nodata&   \nodata &  \nodata&   \nodata&   \nodata \\ 
HD004150  &   3973120 &     A0IV &   73.7&       \nodata &   325& 4          &  139.99 &      2.5 &       56 &    1.041 &       NO &    24.73 &     2.23  &   11.09 &    1.685 &      YES  \\ 
HD011413  &   3973376 &      A1V &   74.8&       \nodata &   600& 3          &   50.73 &     0.91 &    55.75 &    1.034 &       NO &    52.32 &     2.15  &   24.33 &    9.817 &      YES  \\ 
HD011636  &   4228096 &      A5V &   18.3&       \nodata &   620& 5          &  811.52 &     3.28 &   247.41 &    0.961 &       NO &    85.76 &     9.38  &    9.14 &    0.924 &       NO  \\ 
HD014055  &   8796416 &    A1Vnn &   36.1&       \nodata &   300& 6          &  282.66 &     6.63 &    42.63 &    1.445 &      YES &   787.83 &   157.57  &       5 &   36.377 &      YES  \\ 
          &   8796160 & & & & & & & & & & & & & & & \\
HD014228  &   9021952 &   B8IV/V &   47.5&       \nodata &   115& 5          &   185.6 &     1.66 &   111.81 &    0.969 &       NO & $<$41.23 &    13.74  &       3 &    2.033 &      UPL  \\ 
HD014943  &   9940480 &      A5V &   61.3&       \nodata &   850& 22         &   47.35 &     0.47 &   100.74 &    0.973 &       NO &   \nodata&   \nodata &  \nodata&   \nodata&   \nodata \\ 
HD015004  &   3973632 &    A0III &  198.0&       \nodata &   280& 4          &   31.33 &     0.31 &   101.06 &    1.302 &      YES & $<$35.88 &    11.96  &       3 &    13.64 &      UPL  \\ 
HD015008  &   7345408 &    A1.5V &   41.5&       \nodata &   405& 5          &  178.77 &     1.44 &   124.15 &    0.983 &       NO &    21.13 &     0.44  &   48.02 &     1.07 &       NO  \\ 
HD015646  &   3973888 &      A0V &  118.1&       \nodata &   260& 4          &   19.59 &      0.2 &    97.95 &    1.008 &       NO & $<$10.25 &     3.42  &       3 &    4.875 &      UPL  \\ 
HD016970  &   3974144 &      A3V &   25.1&    Ursa Major &   300& 7          &  353.27 &     3.06 &   115.45 &    1.046 &       NO &    45.29 &     4.75  &    9.53 &    1.223 &       NO  \\ 
HD017254  &  11783424 &      A2V &  124.1&       \nodata &   650& 22         &   29.24 &     0.45 &    64.98 &    0.898 &       NO &   \nodata&   \nodata &  \nodata&   \nodata&   \nodata \\ 
HD018978  &   8794368 &      A4V &   26.4&       \nodata &   350& 5          &  240.39 &     2.63 &     91.4 &     1.03 &       NO & $<$26.56 &     8.86  &       3 &    1.054 &      UPL  \\ 
          &   8794112 & & & & & & & & & & & & & & & \\
HD019356  &   3974656 &      B8V &   28.5&       \nodata &   300& 8,5        & 1404.88 &     6.84 &   205.39 &    1.175 &      YES &   206.23 &    11.79  &   17.49 &    1.586 &      YES  \\ 
HD020315  &   3975168 &      B8V &  197.6&  $\alpha$ Per &    80& 8          &   37.46 &    10.66 &     3.51 &    0.951 &       NO &$<$135.93 &    45.31  &       3 &   32.257 &      UPL  \\ 
HD020888  &  11783680 &      A3V &   58.0&       \nodata &   300& 22         &   36.06 &     0.73 &     49.4 &      0.9 &       NO &   \nodata&   \nodata &  \nodata&   \nodata&   \nodata \\ 
HD021362$^{\ddagger}$  &   3976192 &     B6Vn &  169.8&  $\alpha$ Per &    80& 8          &  313.31 &     3.66 &     85.6 &    8.201 &      YES &    90.93 &     5.74  &   15.84 &    22.25 &      YES  \\ 
HD021551  &   3976448 &      B8V &  266.7&  $\alpha$ Per &    80& 8          &   27.61 &     1.04 &    26.55 &    0.958 &       NO & $<$61.37 &    20.46  &       3 &   20.103 &      UPL  \\ 
HD021981  &   8812544 &      A1V &  113.5&       \nodata &   265& 5          &   39.15 &      0.8 &    48.94 &     0.86 &       NO &   \nodata&   \nodata &  \nodata&   \nodata&   \nodata \\ 
HD023267  &   3976960 &      A0V &  136.4&  $\alpha$ Per &    80& 8          &   35.55 &     0.42 &    84.64 &    2.964 &      YES & $<$55.72 &    18.57  &       3 &   42.209 &      UPL  \\ 
HD023642  &   3977984 &      A0V &  110.4&     Pleiades  &   125& 8          &    17.3 &     1.68 &     10.3 &    1.041 &       NO &$<$237.24 &    79.08  &       3 &  131.549 &      UPL  \\ 
HD023753  &   3978240 &      B8V &  103.7&     Pleiades  &   125& 8          &   39.88 &     2.53 &    15.75 &    0.989 &       NO & $<$83.47 &    27.82  &       3 &   18.786 &      UPL  \\ 
HD023763  &   3978496 &      A1V &  144.9&     Pleiades  &   125& 8          &   18.75 &     1.91 &     9.82 &    1.051 &       NO &$<$128.04 &    42.68  &       3 &   64.946 &      UPL  \\ 
HD023862  &   3978752 & B8IVevar &  118.8&     Pleiades  &   125& 8          &  662.14 &      6.4 &   103.46 &    8.986 &      YES &   202.43 &    23.37  &    8.66 &   24.965 &      YES  \\ 
HD023923  &   3979264 &      B8V &  116.6&     Pleiades  &   125& 8          &   46.21 &     0.89 &    51.92 &    1.749 &      YES &$<$118.80 &     39.6  &       3 &   40.679 &      UPL  \\ 
HD023964  &   3975936 &      A0V &  158.7&     Pleiades  &   125& 8          &   17.15 &     0.25 &     68.6 &    0.974 &       NO &  $<$8.00 &     2.67  &       3 &     4.23 &      UPL  \\ 
HD025860  &  15421440 &   A4.5IV &  132.8&       \nodata &   400& 22         &   26.26 &      0.1 &    262.6 &    0.984 &       NO &   \nodata&   \nodata &  \nodata&   \nodata&   \nodata \\ 
HD026321  &   3979520 &      A0V &  175.1&  $\alpha$ Per  &    80& 8          &   13.48 &     0.11 &   122.55 &    0.976 &       NO & $<$12.2 &     4.07  &       3 &    7.945 &      UPL  \\ 
HD027045  &   3979776 &      A3m &   28.7&       \nodata &   193& 8,5        &  133.49 &     0.87 &   153.44 &    1.105 &      YES &    45.73 &     7.03  &     6.5 &    3.406 &      YES  \\ 
HD027628  &   3980032 &      A3m &   45.7&       Hyades  &   625& 9        &   71.59 &     0.61 &   117.36 &    1.001 &       NO & $<$14.60 &     4.87  &       3 &    1.832 &      UPL  \\ 
HD027749  &   3980288 &      A1m &   47.2&       Hyades  &   625& 9          &   71.68 &     0.58 &   123.59 &    0.982 &       NO & $<$14.39 &      4.8  &       3 &    1.778 &      UPL  \\ 
HD027962  &   3980544 &     A2IV &   45.4&       Hyades  &   625& 9,5        &  149.06 &     2.75 &     54.2 &     0.96 &       NO & $<$96.41 &    32.14  &       3 &     5.61 &      UPL  \\ 
HD028226  &   3980800 &       Am &   48.0&       Hyades  &   625& 9          &   72.96 &     1.02 &    71.53 &    1.051 &       NO &   107.55 &    10.11  &   10.64 &   14.311 &      YES  \\ 
HD028355  &   3981056 &      A7V &   49.2&       Hyades  &   625& 9,5        &  137.19 &      1.1 &   124.72 &    1.274 &      YES &   178.78 &     5.45  &    32.8 &   15.389 &      YES  \\ 
HD028527  &   3981312 &     A6IV &   44.4&       Hyades  &   625& 9,5        &  120.07 &     1.52 &    78.99 &    1.019 &       NO &    37.36 &     5.94  &    3.83 &     2.94 &      YES  \\ 
HD028546  &   3981568 &       Am &   44.3&       Hyades  &   625& 9          &   75.31 &     0.46 &   163.72 &    1.026 &       NO & $<$15.38 &     5.13  &       3 &    1.947 &      UPL  \\ 
HD029388  &   3981824 &      A6V &   45.9&       Hyades  &   625& 9,5        &  183.72 &     1.28 &   143.53 &    0.985 &       NO & $<$21.88 &     7.29  &       3 &    1.087 &      UPL  \\ 
HD029488  &   3982080 &     A5Vn &   48.8&       Hyades  &   625& 9          &  139.51 &     0.51 &   273.55 &    0.987 &       NO & $<$13.36 &     4.45  &       3 &    0.875 &      UPL  \\ 
HD030210  &   3982336 &       Am &   81.7&       Hyades  &   625& 9          &   70.33 &     0.73 &    96.34 &    0.986 &       NO & $<$14.19 &     4.72  &    3.01 &     1.85 &      UPL  \\ 
HD030422  &   3982592 &     A3IV &   57.5&       \nodata &    10& 3,5        &   44.67 &     0.64 &     69.8 &    1.202 &      YES &    64.52 &     1.01  &   63.88 &   16.024 &      YES  \\ 
\enddata
\end{deluxetable} 

\begin{deluxetable}{lcrrcrlrrrrcrrrrc}
\tablewidth{0pt} 
\setlength{\tabcolsep}{0.02in}
\tablenum{1}
\tablecaption{Sample of Stars and MIPS Measurements -- Continue.}
\tablecolumns{19}
\tablehead{
\colhead{Name} & \colhead{AOR Key} & \colhead{SpType} &  \colhead{distance} & \colhead{Cluster/} & \colhead{Age}  &\colhead{Age ref.} &\colhead{F$_{m,24}$}&\colhead{$\sigma_{24}$}&\colhead{SN$_{24}$}&\colhead{R$_{24}$}&\colhead{IRE$_{24}$}& \colhead{F$_{m,70}$}&\colhead{$\sigma_{70}$}&\colhead{SN$_{70}$}&\colhead{R$_{70}$}&\colhead{IRE$_{70}$}
\\
\colhead{} & \colhead{} & \colhead{} &  \colhead{pc} & \colhead{Moving Group} & \colhead{Myr}  &\colhead{} &\colhead{mJy}&\colhead{mJy}&\colhead{}&\colhead{}&\colhead{}& \colhead{mJy}&\colhead{mJy}&\colhead{}&\colhead{}&\colhead{}
}
\startdata
HD031295  &   3982848 &      A0V &   37.0&       \nodata &    10& 3,5        &  167.25 &     1.94 &    86.21 &     1.25 &      YES &   418.72 &     4.64  &   90.24 &   28.952 &      YES  \\ 
HD033254  &   3983104 &      A2m &   53.9&       Hyades  &   625& 9          &   72.62 &     1.39 &    52.24 &        1 &       NO &    21.46 &     2.17  &    9.89 &    2.752 &      YES  \\ 
HD034868  &   3983360 &      A0V &  136.8&       \nodata &   300& 4          &    25.9 &      0.3 &    86.33 &    0.983 &       NO & $<$21.68 &     7.23  &       3 &    7.658 &      UPL  \\ 
HD038056  &   3983616 &      A0V &  132.5&       \nodata &   250& 4,5        &   36.48 &     0.37 &    98.59 &    1.823 &      YES &    49.67 &     1.06  &   46.86 &   22.466 &      YES  \\ 
HD038206  &   3983872 &      A0V &   69.2&       \nodata &     9& 4          &  106.92 &     1.58 &    67.67 &    3.361 &      YES &   342.27 &    12.87  &   26.59 &  100.367 &      YES  \\ 
HD038545  &   3984128 &     A3Vn &  129.5&       \nodata &    13& 4          &   45.89 &     0.72 &    63.74 &    0.979 &       NO & $<$13.6 &     4.53  &       3 &    2.678 &      UPL  \\ 
HD038678  &   8792832 &   A2Vann &   21.5&       \nodata &   231& 8,5        &   860.2 &     17.2 &    50.01 &    2.598 &      YES &   246.62 &    24.66  &      10 &    6.872 &      YES  \\ 
          &   9798208 & & & & & & & & & & & & & & & \\
HD039060  &   8970240 &      A5V &   19.3&   $\beta$ Pic  &    12& 23,1,2        &    7276 &      728 &     9.99 &   23.015 &      YES &  12990.4 &   1825.4  &    7.12 &  379.678 &      YES  \\ 
          &  12613632 & & & & & & & & & & & & & & & \\
HD040335  &   9192192 &       A0 &  112.7&       \nodata &     5& 5          &   18.11 &     0.48 &    37.73 &    0.919 &       NO &   \nodata&   \nodata &  \nodata&   \nodata&   \nodata \\ 
HD042525  &  13588224 &      A0V &  101.8&       \nodata &   300& 22         &   33.39 &     0.32 &   104.34 &    0.976 &       NO &   \nodata&   \nodata &  \nodata&   \nodata&   \nodata \\ 
HD043107  &  13476608 &      B8V &   84.9&       \nodata &    80& 5          &   56.18 &      1.1 &    51.07 &    0.961 &       NO &   \nodata&   \nodata &  \nodata&   \nodata&   \nodata \\ 
HD045557  &   3984640 &      A0V &   88.0&       \nodata &    75& 4          &   35.82 &     0.31 &   115.55 &    0.982 &       NO &  $<$9.54 &     3.18  &       3 &    2.454 &      UPL  \\ 
HD046190  &   9662976 &      A0V &   79.0&       \nodata &     5& 5          &   23.86 &     0.05 &    477.2 &    1.126 &      YES &   \nodata&   \nodata &  \nodata&   \nodata&   \nodata \\ 
HD047332  &  11892224 &     A1IV &  343.6&       \nodata &   400& 22         &   16.41 &     0.28 &    58.61 &    1.019 &       NO &   \nodata&   \nodata &  \nodata&   \nodata&   \nodata \\ 
HD048915  &   9458432 &      A0V &    2.6&       \nodata &    70& 8,5        &  \nodata&   \nodata&   \nodata&   \nodata&   \nodata&  2535.26 &    17.81  &  142.35 &    0.926 &       NO  \\ 
HD057336  &  13588736 &     A0IV &  384.6&       \nodata &   400& 10         &   10.05 &     0.18 &    55.83 &    0.957 &       NO &   \nodata&   \nodata &  \nodata&   \nodata&   \nodata \\ 
HD058142  &   7145472 &      A1V &   76.3&       \nodata &   250& 22         &   97.27 &     0.76 &   127.99 &    0.896 &       NO &   \nodata&   \nodata &  \nodata&   \nodata&   \nodata \\ 
HD058647$^{\dagger}$  &   3985152 &     B9IV &  277.0&       \nodata &     1&   Ae/Be    & 2163.32 &    10.55 &   205.05 &  118.166 &      YES &   267.78 &     3.03  &   88.38 &  131.912 &      YES  \\        
HD058715$^{\ddagger}$  &   3985408 &     B8Ve &   52.2&       \nodata &   100& 22,6      & 1469.42 &     7.75 &    189.6 &    3.413 &      YES &   423.04 &    51.02  &    8.29 &     8.91 &      YES  \\ 
HD065517  &  13589248 &   A2.5IV &  105.9&       \nodata &   350& 22         &   14.76 &     0.22 &    67.09 &    1.031 &       NO &   \nodata&   \nodata &  \nodata&   \nodata&   \nodata \\ 
HD069863  &  13589504 &      A2V &   74.2&       \nodata &   650& 5,11       &   81.78 &      3.6 &    22.72 &    1.014 &       NO &   \nodata&   \nodata &  \nodata&   \nodata&   \nodata \\ 
HD071043  &   3985664 &      A0V &   73.1&       \nodata &    11& 4          &   58.21 &     1.44 &    40.42 &    1.836 &      YES &    97.22 &    29.12  &    3.34 &   28.442 &      YES  \\ 
HD071155  &   3985920 &      A0V &   38.3&       \nodata &   169& 8,5        &  302.39 &     4.13 &    73.22 &    1.437 &      YES &   211.74 &     2.76  &   76.72 &    9.087 &      YES  \\ 
HD073210  &   3986432 &      A5V &  196.1&     Praesepe  &   729& 12         &    24.7 &      0.2 &    123.5 &    1.048 &       NO & $<$10.66 &     3.55  &       3 &    4.078 &      UPL  \\ 
HD073666  &   3986688 &      A1V &  174.8&     Praesepe  &   729& 12         &    16.6 &     0.83 &       20 &    0.979 &       NO & $<$5.26 &     1.75  &    3.01 &    2.818 &      UPL  \\ 
HD073731  &   3986944 &      A5m &  168.1&     Praesepe  &   729& 12         &   30.05 &     0.51 &    58.92 &    0.942 &       NO & $<$6.78 &     2.26  &       3 &    1.973 &      UPL  \\ 
HD073819  &   3987200 &     A6Vn &  183.2&     Praesepe  &   729& 12         &   21.03 &     0.19 &   110.68 &    1.008 &       NO & $<$6.13 &     2.04  &       3 &    2.732 &      UPL  \\ 
HD073871  &   3987456 &    A0III &  160.5&     Praesepe  &   729& 12         &   19.26 &     0.09 &      214 &    0.984 &       NO & $<$6.76 &     2.25  &       3 &    3.231 &      UPL  \\ 
HD074956  &   8794880 &      A1V &   24.4&       \nodata &   390& 8,5        & 1396.18 &    25.37 &    55.03 &    1.107 &      YES &$<$180.08 &    60.03  &       3 &    1.315 &      UPL  \\ 
          &   8794624 & & & & & & & & & & & & & & & \\
HD075416  &   3987712 &      B8V &   96.9&   $\eta$ Cha  &     8& 23,13,14     &  122.44 &     1.29 &    94.91 &    3.273 &      YES &     34.5 &     0.93  &    37.1 &    8.381 &      YES   \\ 
HD076644  &   4239360 &      A7V &   14.6&       \nodata &   620& 5          &  629.55 &      4.1 &   153.55 &    1.028 &       NO &    78.54 &     2.28  &   34.45 &    1.178 &       NO  \\ 
HD079108  &   3987968 &      A0V &  115.2&       \nodata &   320& 4          &   44.98 &     1.16 &    38.78 &    1.791 &      YES &    84.35 &     2.52  &   33.47 &   30.357 &      YES  \\ 
HD080007  &  10091008 &     A2IV &   34.1&       \nodata &   260& 5          &  1768.9 &     4.54 &   389.63 &    1.006 &       NO &   212.86 &     0.97  &  219.44 &    1.126 &       NO  \\ 
HD080950  &   3988224 &      A0V &   80.8&       \nodata &    80& 4          &  114.67 &     1.53 &    74.95 &      3.5 &      YES &    59.31 &     0.96  &   61.78 &   16.361 &      YES  \\ 
HD082621  &  12398336 &      A2V &   81.9&       \nodata &   285& 5          &  127.75 &     0.85 &   150.29 &    0.979 &       NO &    14.51 &     0.71  &   20.44 &    1.024 &       NO  \\ 
HD083808  &   4231680 &      A5V &   41.5&       \nodata &   400& 5          &  816.76 &     1.28 &   638.09 &    0.984 &       NO &    94.79 &     1.49  &   63.62 &    1.041 &       NO  \\ 
HD087887  &   3988480 &    A0III &   88.1&       \nodata &   295& 4          &  108.94 &     0.27 &   403.48 &    0.983 &       NO &  $<$9.81 &     3.27  &       3 &    0.811 &      UPL  \\ 
HD087901  &   9661440 &      B7V &   23.8&       \nodata &   140& 5          &  1591.5 &     6.57 &   242.24 &    0.992 &       NO &   179.25 &    35.85  &       5 &    1.034 &       NO  \\ 
HD091375  &   9663232 &      A1V &   79.4&       \nodata &   265& 5          &  102.48 &      2.1 &     48.8 &    1.013 &       NO &   \nodata&   \nodata &  \nodata&   \nodata&   \nodata \\ 
HD092467  &   3988992 &    B9.5V &  141.2&       IC2602  &    50& 15         &   13.32 &      1.4 &     9.51 &    0.994 &       NO & $<$46.44 &    15.48  &       3 &   31.787 &      UPL  \\ 
HD092536  &   3989248 &      B8V &  147.1&       IC2602  &    50& 15         &   46.19 &     0.55 &    83.98 &    2.507 &      YES & $<$60.76 &    20.25  &       3 &   30.078 &      UPL  \\ 

\enddata
\end{deluxetable} 

\begin{deluxetable}{lcrrcrlrrrrcrrrrc}
\tablewidth{0pt} 
\setlength{\tabcolsep}{0.02in}
\tablenum{1}
\tablecaption{Sample of Stars and MIPS Measurements -- Continue.}
\tablecolumns{19}
\tablehead{
\colhead{Name} & \colhead{AOR Key} & \colhead{SpType} &  \colhead{distance} & \colhead{Cluster/} & \colhead{Age}  &\colhead{Age ref.} &\colhead{F$_{m,24}$}&\colhead{$\sigma_{24}$}&\colhead{SN$_{24}$}&\colhead{R$_{24}$}&\colhead{IRE$_{24}$}& \colhead{F$_{m,70}$}&\colhead{$\sigma_{70}$}&\colhead{SN$_{70}$}&\colhead{R$_{70}$}&\colhead{IRE$_{70}$}
\\
\colhead{} & \colhead{} & \colhead{} &  \colhead{pc} & \colhead{Moving Group} & \colhead{Myr}  &\colhead{} &\colhead{mJy}&\colhead{mJy}&\colhead{}&\colhead{}&\colhead{}& \colhead{mJy}&\colhead{mJy}&\colhead{}&\colhead{}&\colhead{}
}
\startdata
HD092715  &   3989504 &    B9.5V &  130.7&       IC2602  &    50& 15         &   12.76 &     0.45 &    28.36 &    1.012 &       NO & $<$40.78 &    13.59  &       3 &   29.305 &      UPL  \\ 
HD092783  &   3989760 &      B9V &  138.7&       IC2602  &    50& 15         &   13.39 &     0.37 &    36.19 &     0.99 &       NO & $<$64.45 &    21.48  &       3 &   43.523 &      UPL  \\ 
HD092845  &   3990016 &      A0V &  185.5&       \nodata &   300& 4          &   40.64 &     0.84 &    48.38 &    0.952 &       NO & $<$11.05 &     3.68  &       3 &    2.392 &      UPL  \\ 
HD093540  &   3990272 &      B6V &  143.1&       IC2602  &    50& 15         &   40.04 &      0.6 &    66.73 &    0.952 &       NO & $<$16.35 &     5.45  &       3 &    3.615 &      UPL  \\ 
HD093549  &   3990528 &     B7IV &  131.6&       IC2602  &    50& 15         &   44.95 &     1.12 &    40.13 &    0.936 &       NO & $<$17.2 &     5.73  &       3 &    3.277 &      UPL  \\ 
HD093738  &   3990784 &    B9.5V &  143.9&       IC2602  &    50& 15         &   23.87 &     1.38 &     17.3 &    1.149 &       NO$^{\sharp}$ & $<$15.51 &     5.17  &       3 &    6.748 &      UPL  \\ 
HD095418  &   7596800 &      A1V &   24.3&    Ursa Major &   300& 7,8,5      &  1026.6 &    14.14 &     72.6 &    1.283 &      YES &   421.13 &    84.23  &       5 &    4.849 &      YES  \\ 
          &   7596288 & & & & & & & & & & & & & & & \\
HD097585  &   3991040 &      A0V &  146.2&       \nodata &   300& 4          &   50.65 &     2.98 &       17 &    0.931 &       NO & $<$35.55 &    11.85  &       3 &    5.904 &      UPL  \\ 
HD097633  &   3991296 &      A2V &   54.5&       \nodata &   550& 5          &   395.1 &     2.48 &   159.31 &    1.187 &      YES &    75.17 &     1.88  &   39.98 &    2.092 &      YES  \\ 
HD101452  &  15247104 &       A2 &  128.2&       \nodata &   250& 22         &   13.05 &     0.12 &   108.75 &    0.977 &       NO &   \nodata&   \nodata &  \nodata&   \nodata&   \nodata \\ 
HD102647  &   4314112 &      A3V &   11.1&       \nodata &    50& 8,5        & 1599.68 &    14.68 &   108.97 &      1.4 &      YES &   676.47 &    45.54  &   14.85 &    5.421 &      YES  \\ 
          &   4313856 & & & & & & & & & & & & & & & \\
HD103287  &   3991552 &      A0V &   25.6&    Ursa Major &   300& 7,8,5      &  788.49 &     7.37 &   106.99 &    0.987 &       NO &    96.06 &     1.73  &   55.53 &    1.118 &       NO  \\ 
HD105805  &   3991808 &     A4Vn &   94.3&         Coma  &   500& 16         &   38.09 &     0.46 &     82.8 &    0.947 &       NO &  $<$8.69 &      2.9  &       3 &    1.963 &      UPL  \\ 
HD106591  &   3992064 &      A3V &   25.0&    Ursa Major &   300& 7,5        &  419.18 &     4.13 &    101.5 &    1.101 &      YES &    65.45 &     3.79  &   17.27 &    1.601 &      YES  \\ 
HD108382  &   3992320 &      A4V &   86.5&         Coma  &   500& 16         &  113.99 &     0.99 &   115.14 &    0.987 &       NO &    26.87 &     1.23  &   21.85 &    2.122 &      YES  \\ 
HD108767  &   3992576 &    B9.5V &   26.9&       \nodata &   260& 5          &  412.69 &     2.44 &   169.14 &    0.992 &       NO &    49.26 &      1.9  &   25.93 &    1.107 &       NO  \\ 
HD108945  &   3992832 &   A2pvar &   95.3&         Coma  &   500& 16         &   55.18 &     0.92 &    59.98 &    0.976 &       NO & $<$27.78 &     9.26  &       3 &    4.574 &      UPL  \\ 
HD109307  &   3993088 &     A4Vm &  106.3&         Coma  &   500& 16         &   26.69 &     0.51 &    52.33 &    0.933 &       NO &  $<$6.98 &     2.33  &       3 &    2.235 &      UPL  \\  
HD110304  &   8811008 &     A1IV &   40.0&       \nodata &   400& 5,11       &  \nodata&   \nodata&   \nodata&   \nodata&   \nodata&   144.58 &      8.6  &   16.81 &    1.396 &       NO  \\ 
HD110411  &   3993344 &      A0V &   36.9&       \nodata &    10& 3          &  139.88 &     1.21 &    115.6 &    1.517 &      YES &   247.95 &     2.22  &  111.69 &   25.002 &      YES  \\ 
HD111786  &   3993600 &    A0III &   60.2&       \nodata &   200& 3          &   66.01 &      1.1 &    60.01 &    1.314 &      YES &    69.86 &     3.65  &   19.14 &    12.55 &      YES  \\ 
HD112185  &   3993856 &      A0p &   24.8&    Ursa Major &   300& 7,5        & 1457.98 &    13.41 &   108.72 &     0.99 &       NO &   178.35 &      4.2  &   42.46 &      1.1 &       NO  \\ 
HD112413  &   3994112 &    A0spe &   33.8&       \nodata &   350& 5          &  337.05 &     7.84 &    42.99 &    0.922 &       NO &    40.27 &     1.75  &   23.01 &    1.032 &       NO  \\ 
HD115892  &   3994368 &      A2V &   18.0&       \nodata &   350& 5          &  683.66 &     2.41 &   283.68 &    1.209 &      YES &    96.62 &     6.47  &   14.93 &    1.593 &      YES  \\ 
HD116706  &   3994624 &     A3IV &   78.0&         Coma  &   500& 16         &   40.88 &     0.65 &    62.89 &    0.945 &       NO & $<$10.25 &     3.42  &       3 &    2.211 &      UPL  \\ 
HD116842  &   3714304 &     A6Vn &   24.9&   Ursa Major  &   300& 7          &258.68   &   1.47   & 175.97   &  0.985   &     NO   &  33.42   &   3.68    &  9.08   &  1.156   &     NO   \\   
HD118878  &   3994880 &      A0V &  121.2&     Upper Cen &    15& 14        &   20.19 &     0.33 &    61.18 &     0.96 &       NO &  $<$7.01 &     2.34  &       3 &    3.051 &      UPL  \\ 
HD123445  &   3995136 &      B9V &  218.8&         UCen  &    15& 14         &   18.96 &     0.54 &    35.11 &    0.963 &       NO & $<$12.91 &        4  &    3.23 &     6.14 &      UPL  \\ 
HD125162  &   3995392 &      A0p &   29.8&       \nodata &   313& 8,5        &   270.8 &     2.32 &   116.72 &    1.384 &      YES &   364.67 &     3.88  &   93.99 &   17.225 &      YES  \\ 
HD126135  &   3995648 &      B8V &  155.5&     Upper Cen &    15& 14         &   22.49 &      1.7 &    13.23 &    2.008 &      YES & $<$34.54 &    11.51  &       3 &   27.998 &      UPL  \\ 
HD126997  &   3995904 &    A0.5V &  144.5&     Upper Cen &    15& 14         &   19.67 &      0.2 &    98.35 &    0.986 &       NO & $<$8.56 &     2.85  &       3 &    3.988 &      UPL  \\ 
HD128207  &   3996160 &      B8V &  128.5&     Upper Cen &    15& 14         &   30.99 &     0.67 &    46.25 &    1.139 &      YES & $<$11.65 &     3.88  &       3 &    3.908 &      UPL  \\ 
HD128998  &  15247360 &      A1V &  131.4&       \nodata &   250& 22         &   32.05 &     0.28 &   114.46 &    0.983 &       NO &   \nodata&   \nodata &  \nodata&   \nodata&   \nodata \\ 
HD129655  &  15421696 &       A2 &  128.5&       \nodata &   300& 22         &   16.53 &     0.12 &   137.75 &    0.999 &       NO &   \nodata&   \nodata &  \nodata&   \nodata&   \nodata \\ 
HD130697  &   3996416 &      A2V &  128.2&     Upper Cen &    15& 14         &   17.49 &     0.44 &    39.75 &    0.933 &       NO &  $<$9.09 &     3.03  &       3 &    4.499 &      UPL  \\ 
HD130841  &   6037760 &     A3IV &   23.7&       \nodata &   495& 5          &  785.59 &     3.59 &   218.83 &    0.979 &       NO &     81.9 &     3.08  &   26.59 &    0.927 &       NO  \\ 
HD132238  &   3996672 &      B8V &  191.9&     Upper Cen &    15& 14         &    30.6 &     0.91 &    33.63 &    1.977 &      YES & $<$12.81 &     4.27  &       3 &    7.601 &      UPL  \\ 
HD133880  &   3996928 &  B8IV... &  126.6&     Upper Cen &    15& 14         &   27.26 &     0.41 &    66.49 &    0.973 &       NO & $<$12.81 &     4.27  &       3 &    4.293 &      UPL  \\ 
HD133937  &   3997184 &      B7V &  136.2&     Upper Cen &    15& 14         &   23.09 &     0.23 &   100.39 &    0.959 &       NO & $<$11.25 &     3.75  &       3 &    4.332 &      UPL  \\ 
HD135454  &   3997440 &      B9V &  137.2&     Upper Cen &    15& 14         &   17.71 &     0.15 &   118.07 &    1.311 &      YES &  $<$8.00 &     2.67  &       3 &    5.502 &      UPL  \\ 
HD135742  &   3997696 &      B8V &   49.1&       \nodata &   100& 8,5        &  501.99 &     1.68 &    298.8 &    0.988 &       NO &     71.5 &     8.37  &    8.54 &    1.322 &       NO  \\ 
HD136246  &   3997952 &      A1V &  143.5&     Upper Cen &    15& 14         &   14.93 &     0.22 &    67.86 &    1.288 &      YES &    34.35 &     1.41  &   24.36 &   27.614 &      YES  \\ 
HD136347  &   3998208 &     A0sp &  123.5&     Upper Cen &    15& 14         &   15.91 &     2.67 &     5.96 &    1.009 &       NO &  $<$8.35 &     2.78  &       3 &    4.925 &      UPL  \\ 
\enddata
\end{deluxetable} 

\begin{deluxetable}{lcrrcrlrrrrcrrrrc}
  \tablewidth{0pt} \setlength{\tabcolsep}{0.02in} \tablenum{1}
  \tablecaption{Sample of Stars and MIPS Measurements -- Continue.}
  \tablecolumns{19} \tablehead{ \colhead{Name} & \colhead{AOR Key} &
    \colhead{SpType} & \colhead{distance} & \colhead{Cluster/} &
    \colhead{Age} &\colhead{Age ref.}
    &\colhead{F$_{m,24}$}&\colhead{$\sigma_{24}$}&\colhead{SN$_{24}$}&\colhead{R$_{24}$}&\colhead{IRE$_{24}$}&
    \colhead{F$_{m,70}$}&\colhead{$\sigma_{70}$}&\colhead{SN$_{70}$}&\colhead{R$_{70}$}&\colhead{IRE$_{70}$}
    \\
    \colhead{} & \colhead{} & \colhead{} & \colhead{pc} &
    \colhead{Moving Group} & \colhead{Myr} &\colhead{}
    &\colhead{mJy}&\colhead{mJy}&\colhead{}&\colhead{}&\colhead{}&
    \colhead{mJy}&\colhead{mJy}&\colhead{}&\colhead{}&\colhead{} }
  \startdata
HD136482  &   3998464 &    B8.5V &  124.5&     Upper Cen &    15& 14         &   42.11 &      0.3 &   140.37 &    2.966 &      YES & $<$20.91 &     6.97  &       3 &   13.485 &      UPL  \\ 
HD137015  &   3998720 &      A2V &  146.6&     Upper Cen &    15& 14         &   15.18 &     0.26 &    58.38 &    1.183 &      YES & $<$18.01 &        6  &       3 &   13.068 &      UPL  \\ 
HD137919  &   3998976 &      B9V &  154.3&     Upper Cen &    15& 14         &   21.04 &      6.1 &     3.45 &    1.141 &       NO$^{\sharp}$ & $<$15.31 &      5.1  &       3 &     7.76 &      UPL  \\ 
HD138923  &   3999232 &      B8V &  112.5&     Upper Cen &    15& 14         &   58.05 &     0.37 &   156.89 &    2.854 &      YES & $<$17.21 &     5.74  &       3 &    7.675 &      UPL  \\ 
HD139006  &   8793600 &      A0V &   22.9&       \nodata &   314& 8,5        & 1261.63 &    15.46 &    81.61 &    1.436 &      YES &   542.03 &     80.7  &    6.72 &    5.633 &      YES  \\ 
          &   8793344 & & & & & & & & & & & & & & & \\
HD139160  &   3999488 &     B9IV &  184.2&     Upper Sco &     5& 14         &   20.98 &      0.1 &    209.8 &    0.968 &       NO & $<$35.49 &    11.83  &       3 &   15.048 &      UPL  \\ 
HD142703  &   3999744 &  A2Ib/II &   52.9&       \nodata &   300& 17         &   54.27 &     0.39 &   139.15 &    0.988 &       NO & $<$46.51 &     15.5  &       3 &     7.63 &      UPL  \\ 
HD144661  &   4000256 &   B8IV/V &  117.6&     Upper Sco &     5& 14         &   17.19 &     0.24 &    71.63 &    1.004 &       NO & $<$11.13 &     3.71  &       3 &    6.076 &      UPL  \\ 
HD145964  &   4000768 &      B9V &  105.8&     Upper Sco &     5& 14         &   19.01 &      1.6 &    11.88 &     0.95 &       NO & $<$15.72 &     5.24  &       3 &    7.248 &      UPL  \\ 
HD158460  &   7970048 &      A1V &  104.4&       \nodata &   260& 5          &   53.15 &     0.34 &   156.32 &    1.155 &      YES &    18.26 &     1.19  &   15.34 &    3.646 &      YES  \\ 
HD158485  &   7969280 &      A4V &  109.3&       \nodata &   420& 5          &      24 &      0.1 &      240 &     0.94 &       NO &   \nodata&   \nodata &  \nodata&   \nodata&   \nodata \\ 
HD161868  &   8797184 &      A0V &   29.1&       \nodata &   184& 8,5        &  413.51 &    18.31 &    22.58 &    1.722 &      YES &  1085.19 &      217  &       5 &   40.819 &      YES  \\ 
          &   8796928 & & & & & & & & & & & & & & & \\
HD163466  &  12872448 &       A2 &  196.5&       \nodata &   310& 5          &   20.15 &     0.13 &      155 &    0.958 &       NO &   \nodata&   \nodata &  \nodata&   \nodata&   \nodata \\ 
HD165459  &   9851392 &      A2V &   89.3&       \nodata &     5& 5          &   25.32 &     0.25 &   101.28 &    1.467 &      YES &    25.93 &     1.08  &   24.01 &   13.664 &      YES  \\ 
HD172167  &   9547776 &      A0V &    7.8&       Castor  &   200& 1,2          &  8900   &     89   &    100   &   1.12   &    YES   & 11416.1  &  2283.22  &       5 &   13.236 &      YE   \\ 
          &   9458432 & & & & & & & & & & & & & & & \\
HD172728  &  12872704 &      A0V &  130.5&       \nodata &   210& 5          &   31.33 &     0.15 &   208.87 &    0.918 &       NO &   \nodata&   \nodata &  \nodata&   \nodata&   \nodata \\ 
HD181296  &   8935424 &     A0Vn &   47.7&       Tucana  &    30& 23,18         &  382.43 &     6.46 &     59.2 &    5.186 &      YES &   408.53 &    41.37  &    9.87 &   50.033 &      YES  \\
          &   8935168 & & & & & & & & & & & & & & & \\
HD183324  &   4002048 &      A0V &   59.0&       \nodata &    10& 3          &    50.3 &     1.25 &    40.24 &    1.114 &      YES &    30.76 &     4.89  &    6.29 &     6.29 &      YES  \\
HD188228  &   3725824 &      A0V &   32.5&       \nodata &    10& 8,5          &170.75   &   0.55   & 310.45   &  0.977   &     NO   &  68.95   &    6.1    &  11.3   &  3.695   &    YES   \\
HD193281  &   4002560 &    A2III &  218.3&       \nodata &   600& 3          &   27.26 &     1.93 &    14.12 &     1.01 &       NO &  $<$7.96 &     2.65  &       3 &    2.649 &      UPL  \\
HD198160  &   4002816 & A2.5IV/V &   73.2&       \nodata &   600& 3          &   62.52 &     0.64 &    97.69 &     0.96 &       NO &    62.07 &     5.78  &   10.74 &    8.609 &      YES  \\
HD204041  &   4003328 &     A1IV &   87.3&       \nodata &   400& 3          &   33.54 &     0.37 &    90.65 &    1.124 &      YES &  $<$8.78 &     2.93  &       3 &    2.647 &      UPL  \\
HD209952  &   7979008 &     B7IV &   31.1&       \nodata &   100& 5          &  979.38 &     1.78 &   550.21 &     0.97 &       NO &   120.77 &     0.71  &   170.1 &    1.129 &       NO  \\
          &   7345152 & & & & & & & & & & & & & & & \\
          &   9661696 & & & & & & & & & & & & & & & \\
          &  13642240 & & & & & & & & & & & & & & & \\
HD210049  &   3729920 &      A2V &   40.0&       \nodata &   245& 5            &128.47   &   0.38   & 338.08   &  1.014   &     NO   &$<$37.84   &  12.61    &     3   &  2.775   &    UPL   \\
HD210111  &   4003584 & A2III/IV &   78.7&       \nodata &   700& 3          &   36.59 &     1.25 &    29.27 &    0.979 &       NO &  $<$9.46 &     3.15  &       3 &    2.264 &      UPL  \\
HD210418  &   4003840 &      A2V &   29.6&       \nodata &   450& 5          &  336.71 &     3.32 &   101.42 &    0.994 &       NO &    46.82 &     3.48  &   13.45 &    1.248 &       NO  \\
HD213320  &   4004096 &    A0IVs &   81.4&       \nodata &   300& 4          &   71.85 &     0.85 &    84.53 &    1.005 &       NO & $<$12.24 &     4.08  &       3 &      1.6 &      UPL  \\
HD214923  &   4004352 &      B8V &   63.9&       \nodata &   120& 5          &  241.91 &        2 &   120.96 &    0.977 &       NO &    34.62 &     1.51  &   22.93 &    1.295 &       NO  \\
HD215789  &   4004608 &      A3V &   39.7&       \nodata &   470& 5          &  363.17 &     2.37 &   153.24 &    1.007 &       NO &    45.67 &     4.43  &   10.31 &    1.143 &       NO  \\
HD216956  &   4889088 &      A3V &    7.7&       Castor  &   200& 1,2        &    3850 &      190 &    20.26 &    1.188 &      YES &  9057.11 &    736.4  &    12.3 &   25.189 &      YES  \\ 
          &  12582400 & & & & & & & & & & & & & & & \\
HD216627  &   4004864 &      A3V &   48.9&    Ursa Major &   300& 7,19       &  420.03 &     2.16 &   194.46 &    0.985 &       NO &     54.7 &     7.12  &    7.68 &     1.18 &       NO  \\
HD221756  &   4004608 &    A1III &   71.6&       \nodata &   130& 3          &   62.46 &     1.08 &    57.83 &    1.184 &      YES &    41.52 &     2.68  &   15.49 &    7.283 &      YES  \\
HD225200  &   3972608 &      A0V &  129.0&      Blanco 1 &    90& 20,21      &    34.7 &      0.7 &    49.57 &    1.495 &      YES &   101.05 &     2.36  &   42.82 &   39.857 &      YES  \\
\enddata 
\tablerefs{1: \citet{barrado98}; 2: \citet{song00}; 3:
    \citet{paunzen97}; 4: \citet{gerbaldi99}; 5: \citet{rieke05};
    6: \citet{westin85}; 7: \citet{king03}; 8: \citet{song01}; 9:
    \citet{perryman98}; 10: \citet{corbally96}; 11:
    \citet{corbally84}; 12: \citet{bonatto04}; 13:
    \citet{mamajek00}; 14: \citet{dezeeuw99}; 15: \citet{randich01};
    16: \citet{odenkirchen98}; 17: \citet{iliev95}; 18:
    \cite{zuckerman01}; 19: \citet{eggen83}; 20: \citet{panagi97}; 21:
    \citet{lynga84}; 22: this study; 23: \citet{zuckerman04}.}
\tablecomments{$^{\flat}$ Age is set to 0 for the young stars that are
  below the ZAMS on the HRD; $^{\dagger}$ Herbig Ae/Be star; $^{\ddagger}$ Be stars;
$^{\sharp}$ stars are disregarded having 24 \um excess due to $\mathcal{X}_{24}<$3.}
\end{deluxetable}
\end{turnpage}

\clearpage 

\begin{deluxetable}{ccclccccccc}
\tabletypesize{\scriptsize}
\tablewidth{0pt} 
\tablenum{2}
\tablecaption{Information on the Additional IRAS Stars\label{tab_iras_stars}}
\tablecolumns{11}
\tablehead{
\colhead{Name} & \colhead{2MASS $K_s$} & \colhead{Age} &
\colhead{SpType} & \colhead{F$_{25}$} & \colhead{R$_{25}$} &
\colhead{F$_{60}$} & \colhead{R$_{60}$} & \colhead{$\sigma_{R_{60}}$} &
\colhead{$\mathcal{X}_{60}$} & \colhead{upper limit} \\
\colhead{} & \colhead{mag} & \colhead{Myr} &
\colhead{} & \colhead{Jy} & \colhead{} &
\colhead{Jy} & \colhead{} & \colhead{} &
\colhead{} & \colhead{}
}
\startdata 
HD050241   &     2.62 &   660 &    A7IV  &  0.891 &  0.993 &  0.230 &  1.48 &   0.31 &   1.54 &   2.10   \\
HD005448   &     3.50 &   600 &    A5V   &  0.395 &  0.994 &  0.118 &  1.71 &   0.46 &   1.54 &   2.63   \\
HD033904   &     3.61 &   150 &    B9IV  &  0.355 &  1.032 &  0.127 &  2.13 &   0.43 &   2.65 &   2.98   \\
HD141003   &     3.31 &   300 &    A2IV  &  0.439 &  0.964 &  0.177 &  2.24 &   0.54 &   2.31 &   3.31   \\
HD027376   &     3.91 &   160 &    B9V   &  0.247 &  0.947 &  0.103 &  2.27 &   0.52 &   2.44 &   3.32   \\
HD130109   &     3.72 &   300 &    A0V   &  0.353 &  1.088 &  0.125 &  2.22 &   0.55 &   2.20 &   3.33   \\
HD077327   &     3.42 &   120 &    A1Vne &  0.478 &  1.165 &  0.161 &  2.26 &   0.56 &   2.23 &   3.39   \\
HD080081   &     3.51 &   395 &    A3V   &  0.393 &  1.042 &  0.151 &  2.31 &   0.58 &   2.27 &   3.46   \\
HD056537   &     3.38 &   560 &    A3V   &  0.586 &  1.320 &  0.177 &  2.30 &   0.60 &   2.17 &   3.49   \\
HD118098   &     3.13 &   510 &    A3V   &  0.550 &  0.986 &  0.236 &  2.44 &   0.54 &   2.68 &   3.51   \\
HD079469   &     3.93 &   180 &    B9.5V &  0.332 &  1.296 &  0.109 &  2.45 &   0.61 &   2.37 &   3.68   \\
HD153808   &     3.79 &   220 &    A0V   &  0.289 &  0.991 &  0.139 &  2.74 &   0.58 &   3.03 &   3.90   \\
HD088955   &     3.73 &   300 &    A2V   &  0.356 &  1.154 &  0.148 &  2.76 &   0.77 &   2.28 &   4.31   \\
HD198001   &     3.72 &   240 &    A1.5V &  0.318 &  1.023 &  0.155 &  2.87 &   0.78 &   2.41 &   4.42   \\
HD089021   &     3.37 &   410 &    A2IV  &  0.431 &  0.963 &  0.232 &  2.98 &   0.75 &   2.66 &   4.48   \\
HD142105   &     4.12 &   180 &    A3Vn  &  0.247 &  1.147 &  0.125 &  3.34 &   0.60 &   3.89 &   4.55   \\
HD129246   &     3.64 &   320 &    A3IVn &  0.342 &  1.023 &  0.177 &  3.05 &   0.79 &   2.58 &   4.63   \\
HD140436   &     3.77 &   160 &    B9IV+ &  0.237 &  0.793 &  0.171 &  3.30 &   0.73 &   3.17 &   4.75   \\
HD039014   &     3.70 &   540 &    A7V   &  0.314 &  0.988 &  0.841 &  15.24&   1.68 &   8.49 &   18.59  \\
\enddata
\tablecomments{Reference for age used can be found in
  \citet{rieke05}. $\sigma_{R_{60}}$ is the error on the ratio
  (R$_{60}$) between measured flux and predicted flux based on $K_s-$[24]
  color. $\mathcal{X}_{60}$ is the significant of excess emission at
  60 \mm, i.e., (R$_{60}-$1)/$\sigma_{R_{60}}$, while
  the last column is a 2-$\sigma$ upper limit on R$_{60}$, i.e., R$_{60}+2\sigma_{R_{60}}$.}
\end{deluxetable}

\begin{deluxetable}{crrrrrr}
\tabletypesize{\scriptsize}
\tablewidth{0pt} 
\tablenum{3}
\tablecaption{Debris Disk Properties \label{disk_properties}}
\tablecolumns{7}
\tablehead{
\colhead{Name}&\colhead{F$_{ire,24}$} &\colhead{$\mathcal{X}_{24}$} &
\colhead{F$_{ire,70}$} &\colhead{$\mathcal{X}_{70}$} &\colhead{T$_{[24]-[70]}$}
 &\colhead{$f_d$}
\\ 
\colhead{}&\colhead{mJy}&\colhead{}&\colhead{mJy}&\colhead{}&\colhead{K}&
\colhead{}
}
\startdata
& \multicolumn{6}{c}{\underline{group I -- 24 and 70 \um excess disks}}        \\
HD014055   &   87.06 &     13.1 &   766.18 &      4.9 &       75&  6.71e-05  \\
HD019356   &  209.51 &     30.6 &    76.21 &      6.5 &      219&  3.19e-06  \\
HD002262   &   18.06 &      4.5 &    47.81 &     12.8 &       98&  7.07e-06  \\
HD023862   &  588.46 &     91.9 &   194.32 &      8.3 &      235&  3.14e-04  \\
HD027045   &   12.69 &     14.6 &     32.3 &      4.6 &       99&  1.05e-05  \\
HD028355   &   29.51 &     26.8 &   167.16 &     30.7 &       82&  5.59e-05  \\
HD030422   &     7.5 &     11.7 &    60.49 &     59.9 &       76&  4.86e-05  \\
HD031295   &   33.49 &     17.3 &   404.26 &     87.1 &       71&  4.46e-05  \\
HD038056   &   16.47 &     44.5 &    47.46 &     44.8 &       96&  4.81e-05  \\
HD038206   &   75.11 &     47.5 &   338.86 &     26.3 &       86&  1.36e-04  \\
HD038678   &  529.07 &     30.8 &   210.73 &      8.6 &      206&  9.75e-05  \\
HD039060   & 6959.86 &      9.6 &  12956.2 &      7.1 &      108&  1.42e-03  \\
HD071043   &   26.51 &     18.4 &     93.8 &      3.2 &       91&  4.74e-05  \\
HD071155   &   91.95 &     22.3 &   188.44 &     68.3 &      105&  2.51e-05  \\
HD075416   &   85.03 &     65.9 &    30.38 &     32.7 &      221&  4.74e-05  \\
HD079108   &   19.87 &     17.1 &    81.57 &     32.4 &       88&  4.97e-05  \\
HD080950   &    81.9 &     53.5 &    55.68 &     58.0 &      154&  7.48e-05  \\
HD095418   &   226.4 &     16.0 &   334.27 &      4.0 &      116&  1.27e-05  \\
HD097633   &   62.37 &     25.1 &    39.24 &     20.9 &      160&  7.09e-06  \\
HD102647   &  457.38 &     31.2 &   551.69 &     12.1 &      123&  1.99e-05  \\
HD106591   &   38.37 &      9.3 &    24.58 &      6.5 &      159&  4.95e-06  \\
HD110411   &   47.66 &     39.4 &   238.03 &    107.2 &       84&  3.69e-05  \\
HD111786   &   15.79 &     14.4 &    64.29 &     17.6 &       88&  2.95e-05  \\
HD115892   &  118.24 &     49.1 &    35.96 &      5.6 &      250&  1.13e-05  \\
HD125162   &   75.08 &     32.4 &    343.5 &     88.5 &       86&  5.06e-05  \\
HD136246   &    3.34 &     15.2 &    33.11 &     23.5 &       73&  4.78e-05  \\
HD139006   &  383.07 &     24.8 &   445.81 &      5.5 &      125&  1.43e-05  \\
HD158460   &    7.12 &     20.9 &    13.25 &     11.1 &      108&  5.26e-06  \\
HD161868   &  173.41 &      9.5 &   1058.6 &      4.9 &       81&  7.52e-05  \\
HD165459   &    8.06 &     32.2 &    24.03 &     22.2 &       95&  4.66e-05  \\
HD172167   &  955.95 &     10.7 &  10553.6 &      4.6 &       72&  2.33e-05  \\
HD181296   &  308.69 &     47.8 &   400.36 &      9.7 &      120&  1.97e-04  \\
HD183324   &    5.13 &      4.1 &    25.87 &      5.3 &       84&  1.04e-05  \\
HD216956   &  608.31 &      3.2 &  8697.54 &     11.8 &       70&  6.14e-05  \\
HD221756   &    9.69 &      9.0 &    35.82 &     13.4 &       90&  1.92e-05  \\
HD225200   &   11.49 &     16.4 &    98.51 &     41.7 &       75&  7.95e-05  \\
& & & & & & \\
\tableline
& \multicolumn{6}{c}{\underline{group II -- only-70 \um excess disks}}     \\
HD004150 & \nodata & \nodata &   10.06  &     4.5  &  $<$119  &     $<$2.24e-06 \\
HD011413 & \nodata & \nodata &   46.99  &    21.9  &    $<$66 &     $<$2.54e-05 \\
HD028226 & \nodata & \nodata &  100.03  &     9.9  &    $<$60 &     $<$3.51e-05 \\
HD028527 & \nodata & \nodata &   24.65  &     4.1  &    $<$83 &     $<$6.27e-06 \\
HD033254 & \nodata & \nodata &   13.66  &     6.3  &    $<$93 &     $<$6.07e-06 \\
HD108382 & \nodata & \nodata &   14.21  &    11.6  &    $<$85 &     $<$4.23e-06 \\
HD188228 & \nodata & \nodata &   50.29  &     8.2  &    $<$61 &     $<$1.19e-05 \\
HD198160 & \nodata & \nodata &   54.86  &     9.5  &    $<$61 &     $<$1.34e-05 \\
 & & & & & &  \\
\tableline
& \multicolumn{6}{c}{\underline{group III -- stellar photospheres}}       \\   
HD015008  & \nodata & \nodata & \nodata & \nodata &$<$249 &  $<$1.37e-06  \\
HD016970  & \nodata & \nodata & \nodata & \nodata &$<$114 &  $<$1.49e-06  \\
HD076644  & \nodata & \nodata & \nodata & \nodata &$<$170 &  $<$1.44e-06  \\
HD080007  & \nodata & \nodata & \nodata & \nodata &$<$361 &  $<$1.07e-06  \\
HD082621  & \nodata & \nodata & \nodata & \nodata &$<$141 &  $<$8.84e-07  \\
HD083808  & \nodata & \nodata & \nodata & \nodata &$<$125 &  $<$6.80e-07  \\
HD087901  & \nodata & \nodata & \nodata & \nodata &$<$83  &  $<$4.70e-07  \\
HD103287  & \nodata & \nodata & \nodata & \nodata &$<$322 &  $<$2.41e-06  \\
HD108767  & \nodata & \nodata & \nodata & \nodata &$<$146 &  $<$5.30e-07  \\
HD112185  & \nodata & \nodata & \nodata & \nodata &$<$244 &  $<$1.31e-06  \\
HD112413  & \nodata & \nodata & \nodata & \nodata &$<$341 &  $<$2.94e-06  \\
HD116842  & \nodata & \nodata & \nodata & \nodata &$<$99  &  $<$1.67e-06  \\
HD130841  & \nodata & \nodata & \nodata & \nodata &$<$140 &  $<$8.79e-07  \\
HD135742  & \nodata & \nodata & \nodata & \nodata &$<$84  &  $<$4.10e-07  \\
HD209952  & \nodata & \nodata & \nodata & \nodata &$<$205 &  $<$9.16e-08  \\
HD210418  & \nodata & \nodata & \nodata & \nodata &$<$130 &  $<$1.57e-06  \\
HD214923  & \nodata & \nodata & \nodata & \nodata &$<$147 &  $<$4.21e-07  \\
HD215789  & \nodata & \nodata & \nodata & \nodata &$<$108 &  $<$1.40e-06  \\
HD216627  & \nodata & \nodata & \nodata & \nodata &$<$93 &  $<$1.61e-06  \\
\enddata
\end{deluxetable}

\end{document}